 \documentclass[11pt]{article}

 \pdfoutput=1   

\usepackage{graphicx}
\usepackage{amsmath}

\usepackage{enumitem}

\usepackage[francais,english]{babel}

\usepackage{float}
\usepackage{ifthen}
\usepackage{color}

\usepackage{hyperref}

\usepackage{natbib}

\begin{document}


\title{ {\bf Parameterization of Atmospheric Convection} \\
        (COST-ES0905 book) in R. S. Plant and J.-I. Yano, editors.\\
        {\bf Vol.~2}: Current issues and new theories. {\it {Chapter 22}: 
        Formulations of moist thermodynamics for atmospheric modelling\/}.
         World Scientific, Imperial College Press, October 2015. }

\author{by Pascal Marquet and late Jean-Fran\c{c}ois Geleyn. {\it CNRM. M\'et\'eo-France.}}


\maketitle

\vspace*{-13 mm}
 \section{\underline{Introduction}.}
\index{sec25.1}\label{sec25.intro}


     \subsection{Motivations}
     \index{sec25.1.1}\label{sec25.Motivations}

Internal energy, enthalpy and entropy are the key quantities to study thermodynamic properties of the moist atmosphere, because they correspond to the First (internal energy and enthalpy) and Second (entropy) Laws of thermodynamics.
The aim of this chapter is to search for analytical formulas for the specific values of enthalpy and entropy and for the moist-air mixture composing the atmosphere.

It may seem a little bit surprising to initiate and put forward this kind of research, because one may consider that all the basic moist-air properties are already described in the available textbooks dealing with atmospheric thermodynamics \citep{degroot_mazur62, glansdorff_prigogine71, IG73, dufour_vanmieghem75, bejan88, emanuel94, BA1998, zdunkowski_bott04, ambaum10}.
It is however explained in this chapter that there are some aspects which are not easily understood and some problems which are not easy to solve cleanly in order to compute locally the moist-air enthalpy and entropy.
It is shown in next sections that the computations of these local state functions require the knowledge of reference values of enthalpies and entropies for all species forming the moist-air (basically N$_2$, O$_2$ and H$_2$O), and that the Third Law of thermodynamics can be used to improve the present formulas for moist-air enthalpy and entropy.

The Third Law consists in specifying zero-entropy at $0$~K and for the more stable crystalline states of atmospheric species (solid-$\alpha$ O$_2$, solid-$\alpha$  N$_2$ and Ice-Ih without proton disorder). 
Differently, it is of common use in atmospheric science to specify zero-entropies at higher temperatures (typically at $0^{\circ}$~C), for the dry air and for one of the water species (typically either for the liquid or the vapour state, depending on the authors). 
It is explained in next sections that it is possible to avoid these arbitrary assumptions and to rely on the Third Law, leading to new interesting results concerning the study of local values of moist-air entropy.
The same programme can be applied to the computation of specific values of moist-air enthapy, with a special attention to be paid to determine the kind of enthalpy that must be considered and for which zero-values can be set at $0$~K.

The theoretical aspects of the new Third-Law based formulations for specific values of enthalpy and entropy, with possible applications or them, are published in a series of four papers \citep{marquet11,marquet_geleyn13,marquet14,marquet15}.
This Chapter~25 describes and summarizes the content of these four papers.

The reason why internal energy is not considered in the previous and next paragraphs is that enthalpy is more suitable than internal energy to study flowing fluids like the moist atmosphere.
According to the notations given in Section~\ref{sec25.Symbols}, the use of enthalpy $h$ simplifies the descriptions of the pressure-volume work in open systems, whereas internal energy $e_i$ is of more common use at the laboratory scale and for portions of fluids at rest.
Indeed, the First Law can be expressed in terms of the specific enthalpy defined by $h = e_i + R \: T = e_i + p / \rho$, due to the perfect gas and the Mayer's laws ($p=\rho \: R \: T$ and $c_p = c_v + R$) valid for the moist air definitions of $c_p$, $c_v$ and $R$.
Therefore, internal energy will not be studied in following sections because it can be easily computed -- if needed -- from the enthalpy, since the reverse formulas $e_i = h - R \: T = h - p / \rho$ are valid for the moist-air definitions of the variables.

The interest of computing local values for enthalpy and entropy corresponds to new features which are different from the usual demand of determining the equations for observed variables, like the temperature.
It is explained in next sections that the temperature equation can be indeed expressed in terms of $c_p\: dT/dt$ or $d(c_p\:T)/dt$, with no need to apply the Third Law and to determine the reference values of enthalpies.
Similarly, the present moist-air conserved variables are built by assuming adiabatic evolution of closed systems, i.e isentropic processes with no change in total water content $q_t$.
These assumptions prevent from the need to determine  reference values of enthalpies or entropies.
However, the knowledge of Third-Law based specific values for $h$ and $s$ can allow original direct computations of $h$, $s$, $dh/dt$ and $ds/dt$, without making the hypothesis of constant values of $q_t$ and with a possibility to analyze isenthalpic or isentropic properties for open-system evolution of moist-air, namely for $q_t$ varying in time and in space.

The Chapter is organized as follows.
An overview of moist-air atmospheric thermodynamics is presented in section~\ref{sec25.overview}, where the role of reference values are described in some details.
The moist-air specific entropy and enthalpy are then defined in sections~\ref{sec25.s_values} and \ref{sec25.h_values} starting from the standard values computed by applying Third-Laws.

Some physical properties of the Third-Law moist-air entropy are presented in section~\ref{sec25.Physical_properties_S}, including comparisons with well-known alternative potential temperatures.
Several applications are shown in section~\ref{sec25.Applications_thetas}: isentropic features, transition from stratocumulus to cumulus, top-PBL entrainment and CTEI criterion, turbulence parameterization, squared Brunt-V\"ais\"al\"a Frequency, potential vorticity.
Section~\ref{sec25.Conservative_theta_s} focuses on conservation properties associated with the Third-Law moist-air entropy.

Some physical properties of moist-air enthalpy are described in section~\ref{sec25.Physical_properties_H}.
Comparisons are made with the well-known concept of moist-static energy and possible applications to turbulent surface fluxes are presented.

Final comments are made in the concluding section~\ref{sec25.Conclusion} about possible new applications of moist-air entropy and enthalpy.

 \section{\underline{Overview of atmospheric thermodynamics}.}
\index{sec25.2}\label{sec25.overview}

     \subsection{Moist-air thermodynamics backgrounds}
     \index{sec25.2.1}\label{sec25.Moist_air_thermo}

The results derived in the next sections correspond to the usual hypotheses made in most textbook on moist-air thermodynamics.

The moist-air is made of dry air, water vapour, liquid water and ice species, denoted by the indexes $d$, $v$, $l$ and $i$, respectively.
Cloud liquid or solid species, together with liquid or solid precipitations, are all considered as being at the same temperature as the rest of the parcel (i.e. the one for dry air and water vapour).
It is assumed that the dry-air and water-vapour species are perfect gases.
The same hypothesis of perfect gas is extended to moist air containing cloud droplets and crystals or precipitations, with the condensed phases having zero volume, though impacting on moist values of $c_p$ and $R$.

Due to Dalton's law, the moist-air specific enthalpy and entropy are assumed to be additive functions, leading to the weighted sums equal to
\begin{eqnarray}
h  & = & \, q_d \: h_d 
    \: + \: q_v \: h_v 
    \: + \: q_l \: h_l 
    \: + \: q_i \: h_i 
    \: ,
\label{def_h} \\
s  & = & \, q_d \: s_d 
    \: + \: q_v \: s_v 
    \: + \: q_l \: s_l 
    \: + \: q_i \: s_i 
    \: .
\label{def_s}
\end{eqnarray}
The parcels of fluid are assumed to be small enough to be homogeneous, but large enough to allow the definition of temperature and pressure variables.
The total mass and the density of a parcel are equal to $m$ (in kg) and $\rho$.
The mass and density of the species are denoted by $m_k$ and $\rho_k$ (values of $k$ are ``d'' for dry air,  ``v'' for water vapour,  ``l'' for liquid water and ``i'' for ice).
The specific contents the mixing ratios are equal to $q_k=m_k/m = \rho_k/\rho$ and $r_k=q_k/q_d=\rho_k/\rho_d$.
The total water specific contents and mixing ratios are denoted by $q_t=q_v+q_l+q_i$ and $r_t=q_t/q_d$.
The sum of the specific contents is equal to $q_d + q_t= 1$.

Three equations of state exist for the moist-air, dry-air and water-vapour perfect gases. 
They can be written $p  = \rho \: R \: T$, $p_d  = \rho \; q_d \: R_d \: T$ and $e = \rho \; q_v \: R_v \: T$.
The property $p=p_d+e$ corresponds to $R=q_d \: R_d \: + \: q_v \: R_v$.

It is assumed that the specific heat at constant pressure of vapour species ($c_{pd}$ and $c_{pv}$), liquid ($c_l$) and solid ($c_i$) are constant terms in the atmospheric range of temperatures (from $150$ to $350$~K).
The specific heat of moist air is equal to the weighted sum $c_p=q_d \: c_{pd} \: + \: q_v \: c_{pv} \: + \: q_l \: c_l \: + \: q_i \: c_i$.

The latent heats of fusion, sublimation and vapourisation are created with changes in enthalpy associated with changes of phases.
For water species, $L_{\mathrm{vap}} = h_v - h_l >0$, where $h_v$ and $h_l$ are the specific values computed per unit mass of moist air.
Similarly, $L_{\mathrm{sub}} = h_v - h_i >0$ and $L_{\mathrm{fus}} = h_l - h_i >0$.

The Gibbs' function is equal to $\mu = h - T \: s$.
It depends both on enthalpy, entropy and absolute temperature.
The important property is that Gibbs' functions are the same for two phases in stable equilibrium states (for instance for liquid and vapour states at the saturation equilibrium curve).
This excludes the metastable states (for instance if super-cooled water is considered).
The consequence is that the change in entropy created by a change of phase is equal to the latent heat divided by the equilibrium temperature ($L_{\mathrm{vap}}/T >0$ for the vaporization processes).

Kirchhoff's  Law are derived with the assumption of constant values for the specific heat at constant pressure, leading for instance to  $dL_{\mathrm{vap}} /dT = d(h_v - h_l)/dT = c_{pv}-c_l$ for the latent heat of vaporization.
The same relationships are valid for the fusion and sublimation processes ($dL_{\mathrm{sub}} /dT =  c_{pv}-c_i$ and $dL_{\mathrm{fus}} /dT =  c_l-c_i$).
If standard values of the latent heats are known at a temperature $T_0$, the latent heat at any temperature $T$ can be determined from $L_{\mathrm{vap}}(T) = L_{\mathrm{vap}}(T_0) \: + \: (c_{pv}-c_l)\:(T-T_0)$, with the equivalent relationships for $L_{\mathrm{sub}}(T)$ and $L_{\mathrm{fus}}(T)$.

The Clausius-Clapeyron equations describe the equilibrium curves between liquid-water (or ice) and water vapour.
The water-saturation equilibrium corresponds to $R_v\:T^2\:de_{sw}/e_{sw} = L_{\mathrm{vap}} \: dT$ and the ice-saturation corresponds to $R_v\:T^2\:de_{si}/e_{si} = L_{\mathrm{sub}} \: dT$.

     \subsection{Special features of atmospheric thermodynamics}
     \index{sec25.2.2}\label{sec25.Atmospheric_thermo}

Many specialized quantities associated with the First and Second Laws of thermodynamics  have been derived and are commonly used in the context of atmospheric science, in particular for applications to turbulent and convective processes.
Examples of those atmospheric quantities are the static energies and the potential temperatures.

The Second Law is mainly expressed in meteorology in terms of the potential temperature $\theta$ via the generic relationship  $s =  c_p \: \ln(\theta) + Cste$, with for instance $\theta = T \:(p_0/p_d)^\kappa$ for the dry air limit.
Moist-air generalizations of $\theta$ exist. They are denoted by $\theta_l$, $\theta_e$,  $\theta'_e$ and $\theta'_w$ and they are all associated with the moist-air entropy and with adiabatic (closed system) or pseudo-adiabatic (saturated open system) assumptions.

The liquid-water and equivalent potential temperatures $\theta_l$ and $\theta_e$ defined in \citet{Betts73} or \citet{emanuel94}  are  conserved quantities for adiabatic motion of a closed parcel of atmosphere, and therefore for isentropic processes of closed parcels of moist-air.
The pseudo-adiabatic wet-bulb version $\theta'_w$ computed in \citet{Saunders57} is a conserved quantity for motions of saturated moist-air, provided that all the condensed water is withdrawn by precipitations.
Even if entropy is conserved by adiabatic upward motions, entropy is removed by precipitations, leading to Saunders formulas which are associated with that increase in specific moist-air entropy which balances the loss by precipitations.

Another equivalent potential temperature is deduced from $\theta'_w$.
It is equal to the dry-air value ($\theta$) attained by a saturated parcel at very low pressure, following a path at constant $\theta'_w$ and until all the water vapour is withdrawn by the precipitations.
This pseudo-adiabatic potential temperature will be denoted by $\theta'_e$.
The numerical values of $\theta_e$ and $\theta'_e$ are close to each others.

The links between the moist-air entropy and the conservative features of these potential temperature are thus subject to hypotheses of closed and adiabatic systems or pseudo-adiabatic processes.
They are not valid for the more general atmospheric states of arbitrary varying  water contents.
This is the motivation for the search for a moist-air potential temperature $\theta_s$ which would be valid for open systems and which would be equivalent to the moist-air entropy via $s =  c_{pd} \: \ln(\theta_s) + Cste$.
The possibility to compute $\theta_s$ (and thus $s$) locally would be interesting for allowing new studies of atmospheric energetics (on the basis of only those general assumptions recalled in the section~\ref{sec25.Moist_air_thermo}).

There is another moist-air potential temperature denoted by $\theta_v$.
It is not directly associated with the moist-air entropy.
It is involved in the computation of the buoyancy force and it represent the leading order approximation of the impact of fluctuation of density.
It is only for the dry-air case that all the moist-air potential temperatures $\theta_l$, $\theta_e$ and $\theta'_w$ coincide with $\theta$, including the virtual potential temperature $\theta_v$.

The computations of moist-air enthalpy are usually conducted with the same hypotheses of closed systems as for the derivation of  $\theta_l$ or $\theta_e$.
The moist-air enthalpy is often expressed by $c_{pd}\:T+L_{vap}\: q_v$ or $c_{pd}\:T - L_{vap}\: q_l$, up to some ``constant'' values.
This demonstrates a close link with what are called ``moist static energies'' (MSE).
The MSEs are associated with the moist-air entropy and are defined by quantities like $c_{pd}\:T+L_{vap}\: q_v+\phi$ or $c_{pd}\:T-L_{vap}\: q_l+\phi$.
The potential energy $g \: z$ is added to the moist-air enthalpy to form the quantity called generalized enthalpy $h+\phi$ \citep{ambaum10}.

It is explained in next sections that these MSE's formulas are derived by making additional hypotheses concerning the zero-enthalpy of dry air and water atmospheric species.
This is the reason why the search for a moist-air enthalpy temperature $T_h$ \citep{Derby2004} which would be equivalent to the moist-air enthalpy via $h =  c_{pd} \: T_h + Cste$ could be interesting for allowing new studies of atmospheric energetics (again with only those general assumptions recalled in the section~\ref{sec25.Moist_air_thermo}).

     \subsection{The role and the calculation of reference values}
     \index{sec25.2.3}\label{sec25.Reference_values}

The aim of this section is to explain why it is difficult to compute the moist-air enthalpy and entropy, and why this is due to the necessity to determine the reference (or standard) values of enthalpies and entropies.

It turns out that it is difficult to compute precisely the specific values of enthalpy and entropy for the moist air understood as an open system made of dry air (O$_2$, N$_2$), water vapour and condensed water species (liquid and ice droplets and precipitations).
The open feature corresponds to the fact that different phases of water can be evaporated from (or precipitated to) the surface, or mixed by convective or turbulent processes, or entrained (or detrained) at the edges of clouds.
For these reasons, except in rare conditions, $q_t$ is a constant neither in time nor in space.

Let us examine the consequences of varying $q_t$ on the impact of reference values by analyzing the usual methods employed in atmospheric science to compute the specific enthalpy \citep{IG73,emanuel94,ambaum10}.
For sake of simplicity, the ice content is assumed to be zero.
The definition~(\ref{def_h}) for $h$ can be rewritten with $q_l=q_t-q_v$ or $q_v=q_t-q_l$ as any of
\begin{eqnarray}
h  & = & \, q_d \: h_d 
    \: + \: q_t \: h_v 
    \: + \: q_l \: (h_l-h_v)
    \: , 
\label{def_h2}\\
h  & = & \, q_d \: h_d 
    \: + \: q_t \: h_l 
    \: + \: q_v \: (h_v -h_l)
    \: .
\label{def_h3}
\end{eqnarray}
The next step is to use the properties $q_d = 1 - q_t$ and $L_{\mathrm{vap}} = h_v -h_l$, leading to
\begin{eqnarray}
h  & = & \,  h_d 
    \: + \: q_t \:( h_v - h_d)
    \: - \: q_l \: L_{\mathrm{vap}}
    \: ,
\label{def_h4} \\
h  & = & \, h_d 
    \: + \: q_t \: ( h_l - h_d)
    \: + \: q_v \: L_{\mathrm{vap}}
    \: .
\label{def_h5}
\end{eqnarray}
Since the specific heats at constant pressure are assumed to be independent of $T$, the enthalpies at the temperature $T$ can be expressed from the enthalpies at the reference temperature $T_r$.
The result is $h_d = (h_d)_r + c_{pd}\:(T-T_r)$ for the dry air, with equivalent relationships for water vapour (index $v$) and liquid water (index $l$).
Equations~(\ref{def_h4}) and (\ref{def_h5}) transform into
\begin{eqnarray}
\!\!\!\!\!\!\!\!
h  & = & \,  \left[ \:  c_{pd} + (  c_{pv} -  c_{pd} ) \: q_t  \: \right] \: T
    \: -  \: L_{\mathrm{vap}} \: q_l
     \nonumber \\
\!\!\!\!\!\!\!\!
   &  &  + \: \left[ \:  (h_d)_r -  c_{pd}  \: T_r \: \right] 
            \: +  q_t \left[ \:  (h_v)_r - (h_d)_r \: \right]
            \:-  q_t \: ( c_{pv} - c_{pd} ) \: T_r
    \: ,
\label{def_h6} \\
\!\!\!\!\!\!\!\!
h  & = & \,  \left[ \:  c_{pd} + (  c_l -  c_{pd} ) \: q_t  \: \right] \: T
    \: +  \: L_{\mathrm{vap}} \: q_v
     \nonumber \\
\!\!\!\!\!\!\!\!
   &  &  + \: \left[ \:  (h_d)_r -  c_{pd}  \: T_r \: \right] 
            \:+  q_t \left[ \:  (h_l)_r - (h_d)_r \: \right]
            \:-  q_t \: ( c_l - c_{pd} ) \: T_r
    \: .
\label{def_h7}
\end{eqnarray}

The second lines of (\ref{def_h6}) and (\ref{def_h7}) are usually discarded.
Since the term $[ \: (h_d)_r - c_{pd} \: T_r \: ]$ is a true constant, it can be removed in both (\ref{def_h6}) and (\ref{def_h7}) because it represent a global offset having no physical meaning.
The other terms depending on $q_t$ can also be neglected if $q_t$ is a constant, i.e. for motions of closed parcels of moist air.

However, if $q_t$ is not a constant, the last quantities depending on $( c_l - c_{pd} )$ and $( c_{pv} - c_{pd} )$ must be taken into account as corrections terms to the first lines.
These last terms could be easily managed, since the numerical value of specific heats are known.
However, the other terms $[ \: (h_l)_r - (h_d)_r \: ]$ and $[ \: (h_v)_r - (h_d)_r \: ]$ depending on the reference values are more problematic for varying $q_t$, since these reference values are not known.
The only possibility to discard these terms is to set arbitrarily $(h_v)_r = (h_d)_r$ in (\ref{def_h6}) or $(h_l)_r = (h_d)_r$ in (\ref{def_h7}).
But these arbitrary choices have never been justified in general or atmospheric thermodynamics.

Alternative formulations are usually derived in terms of the enthalpy expressed ``per unit mass of dry air'.
The quantities $k_w$ and $k$ are defined by Eqs.(4.5.5) and (4.5.4) in \citet{emanuel94} and they are called ``liquid water enthalpy'' and ``moist enthalpy'', respectively.
They correspond to $h/q_d$ with $h$ given by (\ref{def_h2})  and (\ref{def_h3}).
They are written in terms of the mixing ratios, leading to $k_w = h_d + r_t \: h_v -   L_{\mathrm{vap}} \: r_l$ and $k = h_d + r_t \: h_l +  L_{\mathrm{vap}} \: r_v$.
Equations~(\ref{def_h6}) and (\ref{def_h7}) are then replaced by 
\begin{eqnarray}
k_w  & = & \,  (  c_{pd} +   c_{pv} \: r_t  ) \: T
    \: -  \: L_{\mathrm{vap}} \: r_l
     \nonumber \\
   &  & \; + \:  \left[ \:  (h_d)_r -  c_{pd}  \: T_r  \: \right] 
           \; + \:       r_t \: \left[ \:  (h_v)_r -  c_{pv}  \: T_r  \: \right]
    \: ,
\label{def_h8} \\
k    & = & \,  (  c_{pd} +   c_l  \: r_t  ) \: T
    \: +  \: L_{\mathrm{vap}} \: r_v
     \nonumber \\
   &  & \; + \:  \left[ \:  (h_d)_r -  c_{pd}  \: T_r  \: \right] 
           \; + \:   r_t \: \left[ \:  (h_l)_r -  c_l  \: T_r  \: \right] 
    \: .
\label{def_h9} 
\end{eqnarray}
The last lines of (\ref{def_h8}) and (\ref{def_h9}) are discarded in \citet{emanuel94}.
This is indeed valid if $q_t$ is a constant term.
But this is not valid for varying $q_t$, except if the arbitrary hypotheses $(h_v)_r =  c_{pv}  \: T_r$ or $(h_l)_r =  c_l \: T_r$ are made.

The conclusion of this section is that, if $q_t$ is not a constant, it is needed to determine the reference values of $(h_d)_r$ for dry air, and $(h_l)_r$ or $(h_v)_r$ for water species, in order to compute the moist-air enthalpy.
Even if there is a link between the water-vapour and the liquid-water enthalpies, via $(h_v)_r = (h_l)_r + L_{\mathrm{vap}}(T_r)$, no such link exist between the dry-air and any of the water reference values.
The same problem is observed in order to compute specific values of the moist-air entropy, with the need to know the reference entropies and with additional problems created by the need to manage the reference partial pressures for the dry air and the water vapour.

     \subsection{The new proposals based on the Third Law}
     \index{sec25.2.4}\label{sec25.Third_Law_proposals}

It is possible to compute the First and Second Laws moist-air specific quantities $h$ and $s$ without making any {\it a priori\/} assumptions of adiabatic, pseudo-adiabatic or closed parcel of fluid, and without making the arbitrary hypotheses described in the previous section concerning the reference values.
The theoretical aspects of the new Third-Law moist-air formulations for $h$ and $s$, with possible applications or them, are published in a series of four articles \citep{marquet11,marquet_geleyn13,marquet14,marquet15}.

Following the old advice given in  \citet{Richardson22} and continuing the method described in \citet{hauf_holler87}, it  is shown in \citet{marquet11} that it is possible to use the Third Law in order to define a moist-air potential temperature denoted by $\theta_s$, with the moist-air entropy written as $s_{ref} + c_{pd} \: \ln(\theta_s)$ where $s_{ref}$ and $c_{pd}$ are two constant.
This new moist-air potential temperature $\theta_s$ generalizes $\theta_l$ and $\theta_e$  to open systems and  becomes an exact measurement of the moist-air entropy, whatever the changes in water contents $q_v$, $q_l$ or $q_i$ may be.

The same method used for defining the moist-air entropy based on the Third Law is used in \citet{marquet15} to express the moist-air enthalpy without making the assumptions of pseudo-adiabatic, adiabatic or zero-enthalpy at temperature different from $0$~K.

Applications of open moist-air features described in the four articles are to stratocumulus (see Chapters~\ref{sec25.Sc_to_Cu} and \ref{sec25.CTEI}), to moist-air turbulence via calculations of the Brunt-V\"ais\"al\"a frequency described in Chapters~\ref{sec25.moist_turb} and \ref{sec25.moist_BVF}, to the definition of a moist-air potential vorticity (see Chapter~\ref{sec25.moist_PV}) or to comparisons with moist-static energies (see Chapter~\ref{sec25.comparisons_h_Th}).

     \subsection{Computation of standard values for entropies and enthalpies}
     \index{sec25.2.5}\label{sec25.Standard_values}

The aim of this section is to compute the standard values of enthalpy and entropy.
They are denoted by the superscript ``0'', leading to $h^0$ and $s^0$, respectively. 
They must be computed for the three main species which compose the moist-air: Nitrogen (N${}_2$), Oxygen (O${}_2$) and Water (H${}_2$O).

It is worthwhile explaining more precisely in what the reference values are different from the standard ones. 

The {\it standard temperature\/}  is commonly set to any of $0\:^{\circ}$C,  $15\:^{\circ}$C,  $20\:^{\circ}$C or $25\:^{\circ}$C, whereas the standard pressure $p_0$ is set to either $1000$~hPa or $1013.25$~hPa.
The standard values are set to $T_0 =   0\:^{\circ}$C and  $p_0= 1000$~hPa for all species in next sections of this Chapter~25.

The {\it reference temperature\/} correspond to a same value $T_r$ for all species.
Differently from the standard definitions $(p_d)_0 = e_0 = 1000$~hPa, the reference partial pressures $(p_d)_r$ for the dry air and $e_r$ for the water vapour are different.
The reference total pressure is equal to $p_r = (p_d)_r + e_r$ and is  set to $p_r = p_0 = 1000$~hPa hereafter.
The water-vapour partial pressure $e_r$ is equal to the saturation pressure value at $T_r$.
It is thus close to $6.11$~hPa for $T_r =0\:^{\circ}$C.
This value is much smaller that $p_0$ and this explains why the reference values of water-vapour entropy will be different from the standard ones.
The dry-air partial pressure is equal to $ (p_d)_r = p_r - e_r \approx 993.89$~hPa.

It is not possible to compute enthalpy and entropy from the ideal-gas properties only, because the ideal-gas assumption ceases to be valid at very low temperature.
Moreover, it is not possible to use the Sakur-Tetrode equation for computing the expressions for the entropies of N${}_2$, O${}_2$ or H${}_2$O, since they are not mono-atomic ideal gases.
It is assumed that the three main gases N${}_2$, O${}_2$ and H${}_2$O are free of chemical reactions, with the consequence that the specific  enthalpies cannot correspond to the concept of ``standard enthalpies of formation or reaction'' (denoted by $\Delta H^0_f$ and $\Delta H^0_r$ in chemical tables for most species).

The general method used in \citet{marquet15} is to take advantage of the fundamental property of $h$ and $s$ to be state functions, and to imagine at least one reversible path which connect the dead state at $T=0$~K and $p_0=1000$~hPa and the standard state at $T_0$ and $p_0$.

The standard entropy $s^0$ of a substance at $T_0$ and $p_0$ is obtained by:
\vspace{-2.5mm}
\begin{itemize}
\item setting to zero the value at $0$~K by virtue of the Third law for the more stable solid state at $0$~K;
\item computing the integral of $c_p(T)/T$ following the reversible path from $0$~K to $T_0$, for the solid(s), the liquid and the vapour;
\item adding the contribution due to the change of states (latent heats) for solid(s)-solid(s), solid-liquid and liquid-vapour transitions.
\end{itemize}
\vspace{-2mm}
The second and third steps must be understood for a series of solid states between $0$~K and up to the fusion point, possibly different from the more stable one at $0$~K\footnote{ 
Almost all atmospheric species (N${}_2$, O${}_2$, Ar, CO${}_2$) are gaseous at  $T_0 = 273.15$~K$\: = 0\,{}^{\circ}$C.
Only water is different, due to the unusual high fusion point temperature.
More precisely, there are 3 solid states for O$_2$ ($\alpha$, $\beta$ and $\gamma$), 2 solid states for N$_2$ ($\alpha$ and $\beta$) and only one for Ice-Ih (hexagonal ice).}.

Similarly, the thermal part of enthalpy is a thermodynamic state function which is fully determined by the integral of $c_p(T)$ following the same reversible path as the one described for the computations of the entropy.
The thermal enthalpies for N${}_2$, O${}_2$ and H${}_2$O substances are generated by the variations of $c_p(T)$ with $T$ corresponding to progressive excitations of the translation, rotation and vibrational states of the molecules, and by the possible changes of phase represented by the latent heats (but with no or very small impact of changes of pressure).

\begin{figure}[ht]
\centering
\includegraphics[width=0.65\linewidth,angle=0,clip=true]{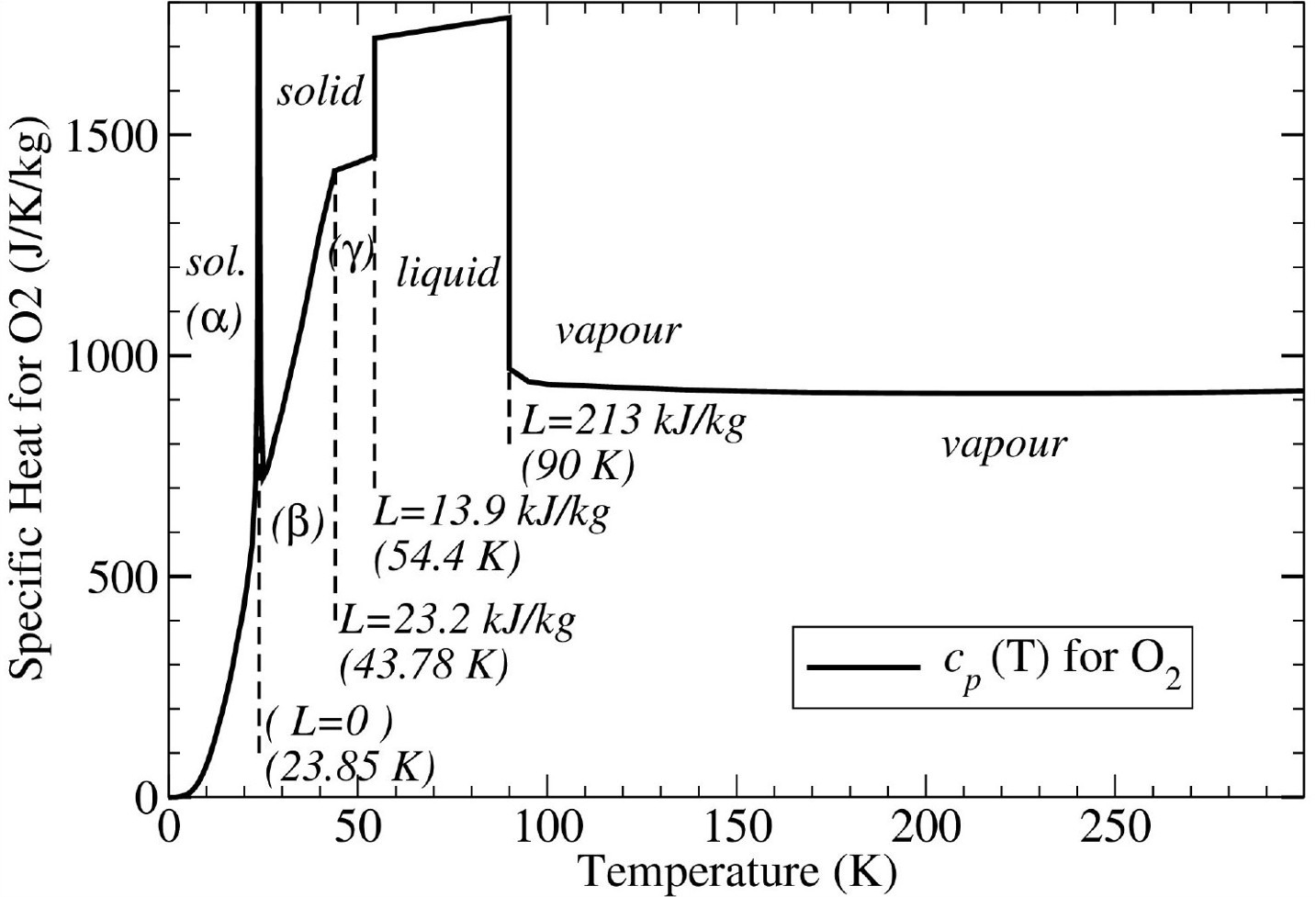}\\
\includegraphics[width=0.65\linewidth,angle=0,clip=true]{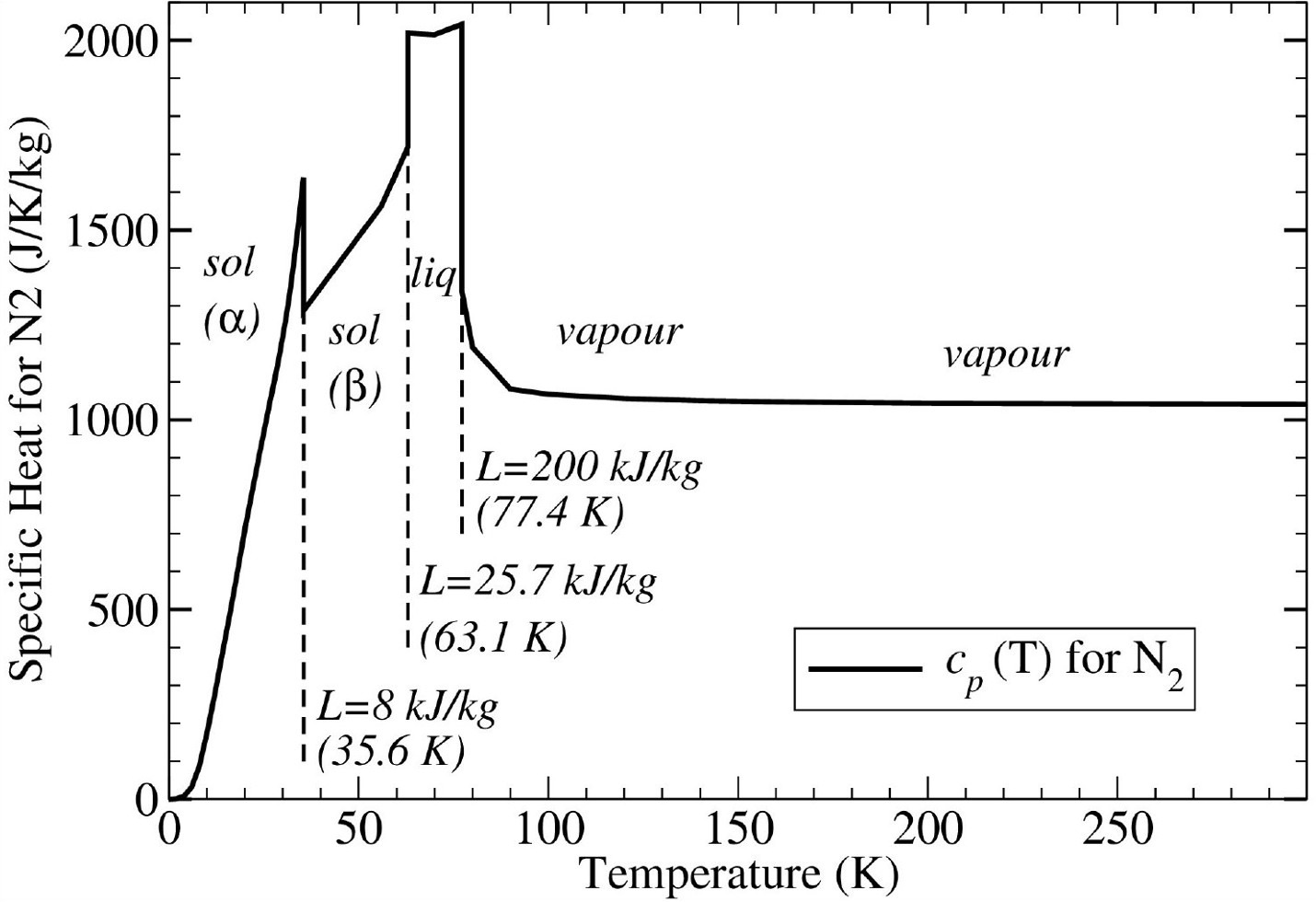}
\caption{\it \small The specific heat $c_p(T)$ at constant pressure ($1000$~hPa) for (a) O${}_2$ and (b) N${}_2$.
Units are J~K${}^{-1}$~kg${}^{-1}$.
There are $3$ solid phases ($\alpha$, $\beta$ and $\gamma$) for O${}_2$ and $2$ solid phases ($\alpha$ and $\beta$) for N${}_2$.
Latent heats are expressed in kJ~kg${}^{-1}$ at each change of phase (solid-solid, solid-liquid, liquid-vapour), together with the corresponding temperature.}
\label{fig25.1}
\end{figure}

The aim is thus to compute integrals of $c_p(T)$ and $c_p(T)/T$, plus the impacts of latent heats existing for N${}_2$, O${}_2$ and H${}_2$O.
Integrals extend from $0$~K to the standard temperature $T_0$ by following a reversible path involving the more stable forms of the substance at each temperature.
These processes are represented by the different terms in the formulas
\begin{align}
  h^{0} \: = \: h(T_0)  & = \: h(0) 
  + \int^{T_0}_{0} \! c_p(T)\:  dT
  + \sum_k L_k
  \: , \label{def_abs_h}
\end{align}
and
\begin{align}
  s^{0} \: = \: s(T_0,p_0)  & = \: s(0,p_0) \; + \, 
 \int^{T_0}_{0} \frac{c_p(T)}{T} \: dT \; + \: \sum_j\:\frac{L_j}{T_j}
  \: . \label{def_abs_s}
\end{align}
The latent heats represented by the $L_j$'s terms are due to changes of phase (solid-solid, melting and vaporization).
Another contribution must be added to Eq.~(\ref{def_abs_s}) if the standard entropy is computed at a pressure different from $1000$~hPa (for instance at the saturating pressure over liquid water or ice).

\begin{figure}[ht]
\centering
\includegraphics[width=0.65\linewidth,angle=0,clip=true]{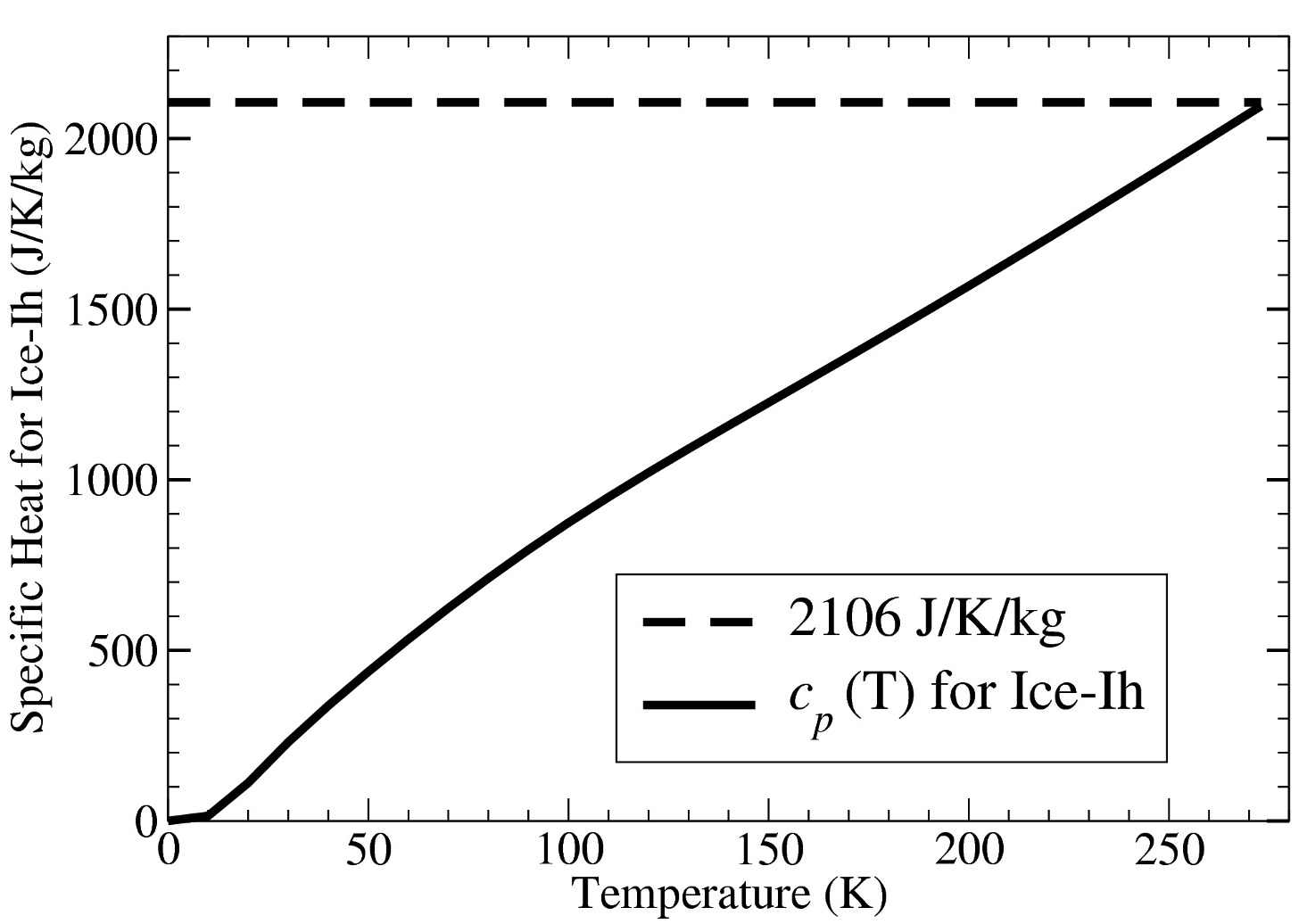}
\caption{\it \small The specific heat $c_p(T)$ at constant pressure ($1000$~hPa) for H${}_2$O (ice-Ih).
Units are J~K${}^{-1}$~kg${}^{-1}$.
The value $2106$~J~K${}^{-1}$~kg${}^{-1}$ is the commonly accepted one in atmospheric science.}
\label{fig25.2}
\end{figure}

The standard values $h^0$ and $s^0$ are computed from Eqs.~(\ref{def_abs_h}) and (\ref{def_abs_s}) and by using the cryogenic datasets described in \citet{marquet15}.
The specific heats and the latent heats for O${}_2$, N${}_2$ and H${}_2$O (ice-Ih) are depicted in Figs.~\ref{fig25.1} and  \ref{fig25.2}.
They are based on the books and papers of \citet{FHH69}, \citet{JPL97}, \citet{MF97}, \citet{KN01}, \citet{FW06} and \citet{Lipinski07}.

\begin{figure}[ht]
\centering
\includegraphics[width=0.9\linewidth,angle=0,clip=true]{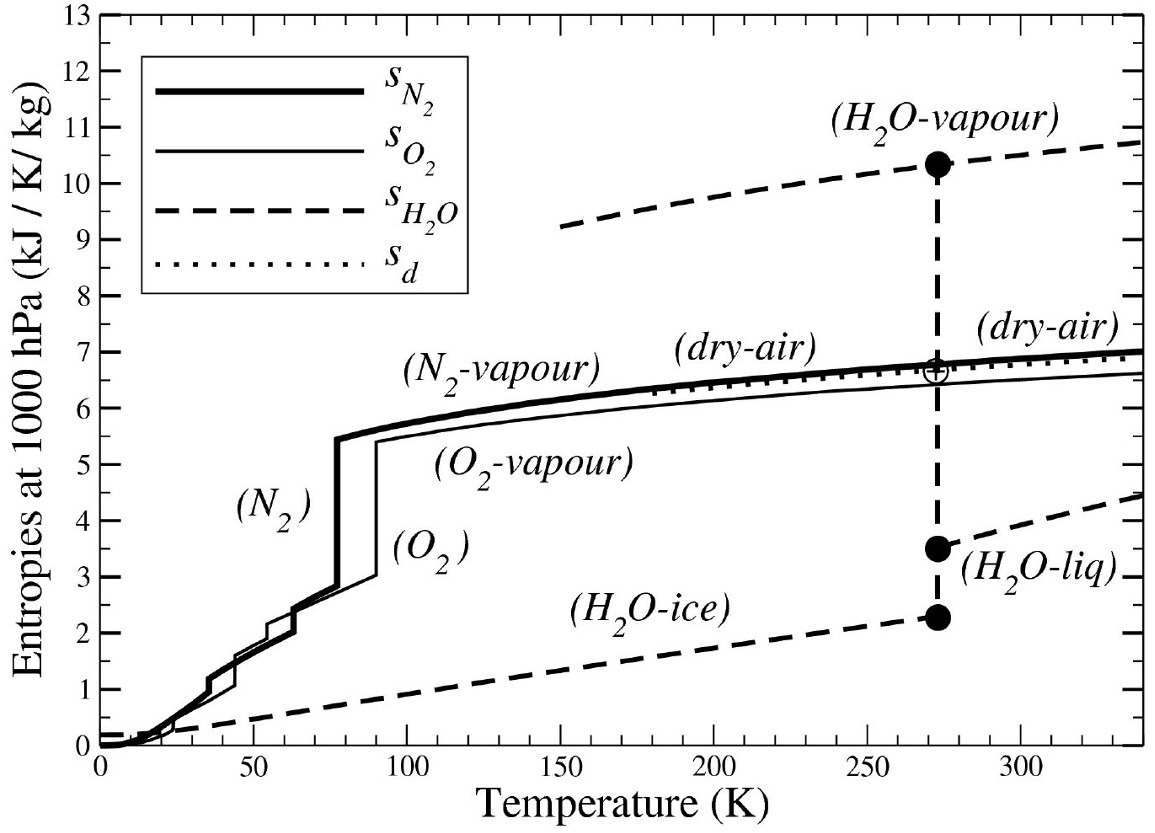}
\caption{\it \small  The entropy diagram at $1000$~hPa for O${}_2$, N${}_2$, dry-air and the three phases of H${}_2$O (ice, liquid, vapour).
Units are kJ~K${}^{-1}$~kg${}^{-1}$.}
\label{fig25.3}
\end{figure}

The Third law means that the entropy of a perfect crystal approaches zero as temperature approaches absolute zero, leading to $s(0,p_0)=0$.
The specific entropies at the pressure $1000$~hPa are shown in Fig.~\ref{fig25.3} for temperatures between $0$ and $340$~K.
The standard entropies computed at $T_0=273.15$~K$\;=0\:^{\circ}$C and $p_0=1000$~hPa are given in Eqs.\ref{def_s0d} to \ref{def_s0i}.
\begin{align}
   s^0_d  & \; \approx \: 6846 \:[\: 6775 \:]\: \mbox{~J}\mbox{~K}^{-1} \mbox{~kg}^{-1}
   \: , \label{def_s0d} \\
   s^0_v  & \; \approx \: 10327 \:[\: 10320 \:]\: \mbox{~J}\mbox{~K}^{-1} \mbox{~kg}^{-1}
   \: , \label{def_s0v} \\
   s^0_l   & \; \approx \: 3520 \:[\: 3517 \:]\: \mbox{~J}\mbox{~K}^{-1} \mbox{~kg}^{-1}
   \: , \label{def_s0l} \\
   s^0_i   & \; \approx \: 2298 \:[\: 2296 \:]\: \mbox{~J}\mbox{~K}^{-1} \mbox{~kg}^{-1}
   \: . \label{def_s0i}
\end{align}
These values of standard entropies are close to those published in Thermodynamic Tables (recalled in bracketed terms in Eqs.\ref{def_s0d} to \ref{def_s0i}) and retained in \citet{hauf_holler87} and \citet{marquet11}.
It is worthwhile noticing that a residual entropy of about $189$~J~K${}^{-1}$~kg${}^{-1}$ must be taken into account for H$_2$O at $0$~K (see the vertical shift for H${}_2$O (ice-Ih) in Fig.~\ref{fig25.3}), due to proton disorder and to remaining randomness of hydrogen bonds at $0$~K \citep{Pauling35,Nagle66}.

\begin{figure}[ht]
\centering
\includegraphics[width=0.67\linewidth,angle=0,clip=true]{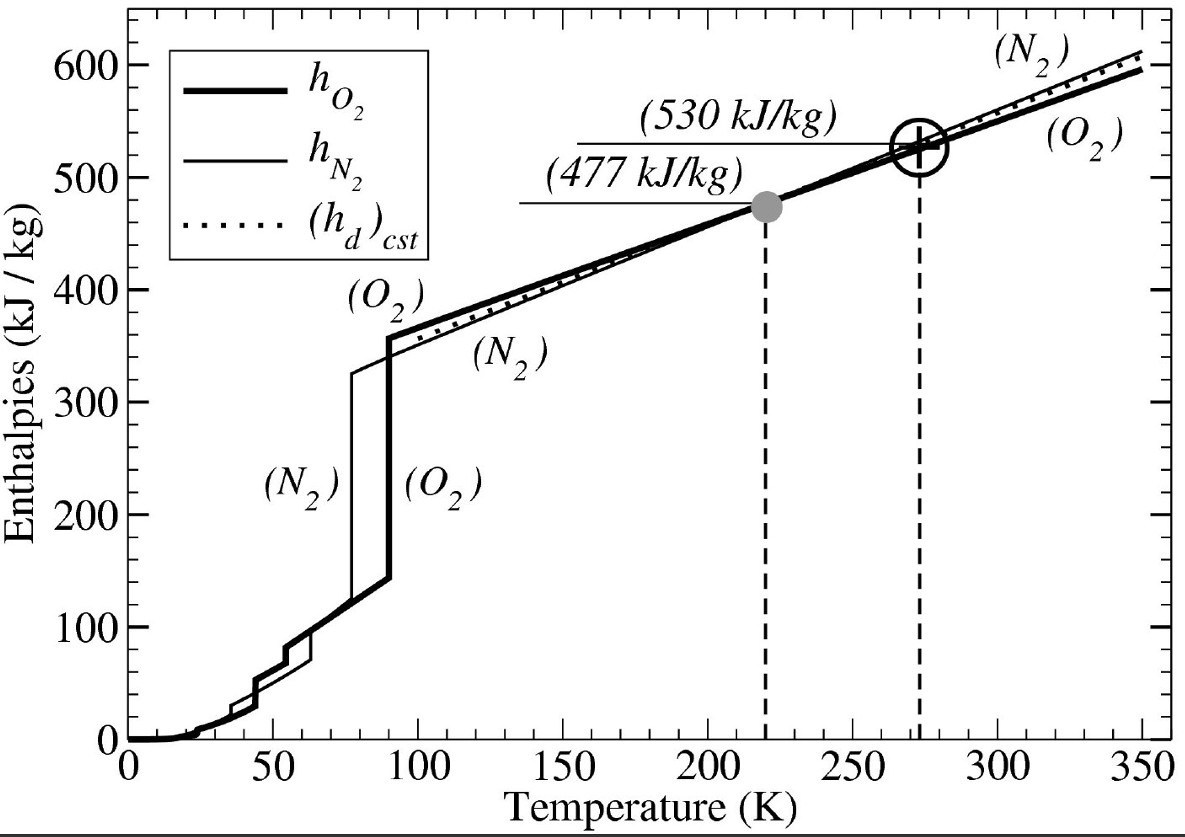}\\
\includegraphics[width=0.67\linewidth,angle=0,clip=true]{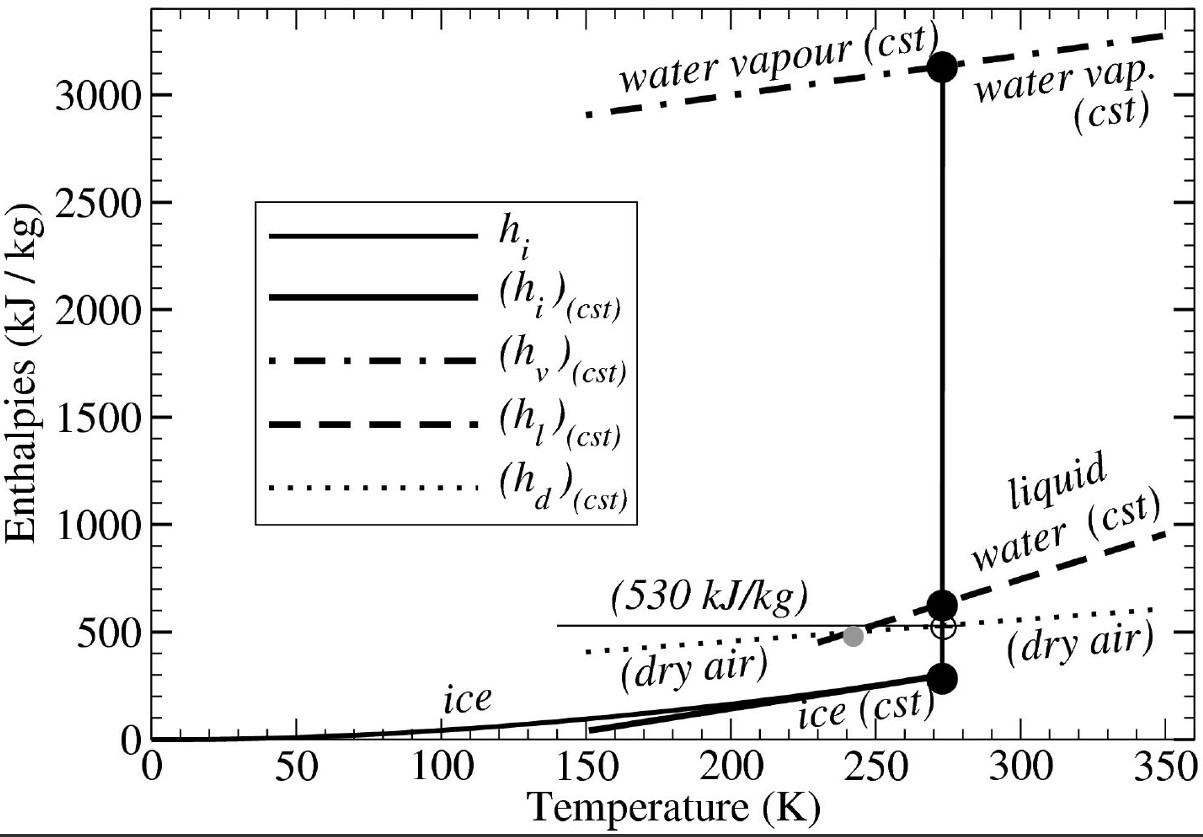}
\caption{\it \small  The enthalpy diagram at $1000$~hPa for (a) O${}_2$, N${}_2$ and dry-air and (b) dry-air and the three phases of H${}_2$O (ice, liquid, vapour).
Units are kJ~kg${}^{-1}$.
The enthalpies for N${}_2$ and O${}_2$ are equal for $T \approx 220$~K in (a) and those for dry air and water vapour are equal for $T \approx 241.4$~K in (b). 
Grey spots label these locations.
The circle-cross in (a) and (b) represent the dry-air value at $0 \:^{\circ}$C ($530$~kJ~kg${}^{-1}$).
}
\label{fig25.4}
\end{figure}

The ice, liquid-water and water-vapour values are $2$, $3$, and $7$ unit higher than the one retained in \citet{hauf_holler87}.
This is in good agreement with the accepted values, with an accuracy better than $0.1$\%, and this provide a validation of the cryogenic datasets described in \citet{marquet15} and depicted in Fig.\ref{fig25.2}.

The dry-air value of standard specific entropies is $71$ unit  higher than the one retained in \citet{hauf_holler87}.
The accuracy is thus coarser for the dry air ($1$\%) than for the water species ($0.1$\%) and the higher uncertainty for the dry-air value is mainly due to the dataset for Oxygen ($0.5$\% for N${}_2$ and $1.3$\% for O${}_2$).
This may be explained by the solid $\alpha$-$\beta$ transition of O${}_2$ occurring at $23.85$~K.
It is a second order transition, with no latent heat associated to it, but with infinite values of $c_p$ leading to a kind of Dirac function for $c_p(T)$, partly visible on Fig.~\ref{fig25.1}(a) (up to $1800$~J~K${}^{-1}$~kg${}^{-1}$ only).
This second order transition is not always taken into account in published values of standard entropy for  O${}_2$.
If a smoother transition were introduced in the cryogenic datasets, the new computed value for $s^0_d$ would be $3$\% lower, making the value retained in \citet{hauf_holler87} compatible with the present one.

The same cryogenic datasets is used to compute dry-air and water-species standard values of specific thermal enthalpies.
It is assumed that a kind of Third law may be use to set the thermal part of the enthalpy to zero at $0$~K.
The specific thermal enthalpies are shown in Fig.\ref{fig25.4} for temperatures from $0$ to $360$~K.
The ({\it cst\/}) curves for water vapour, liquid water and ice are plotted in Fig.~\ref{fig25.4}(b) from the values of enthalpy at $0 \:^{\circ}$C and by assuming that the specific heat at constant pressure is a constant (this explains the difference with the true enthalpy for ice below $200$~K).

The standard values computed at $T_0=273.15$~K$\;=0\:^{\circ}$C are equal to
\begin{align}
   h^0_d  & \;  \approx \: 530  \mbox{~kJ} \mbox{~kg}^{-1}
   \: , \label{def_h0d} \\
   h^0_v  & \;  \approx \: 3133  \mbox{~kJ} \mbox{~kg}^{-1}
   \: , \label{def_h0v} \\
   h^0_l   & \;  \approx \: 632  \mbox{~kJ} \mbox{~kg}^{-1}
   \: , \label{def_h0l} \\
   h^0_i   & \;  \approx \: 298  \mbox{~kJ} \mbox{~kg}^{-1}
   \: . \label{def_h0i}
\end{align}
The value of $h^0_i$ is in close agreement with the value $298.35$~kJ~kg${}^{-1}$ published in \citet{FW06}.
The standard values given in Eqs.~(\ref{def_h0d})-(\ref{def_h0i}) are computed from cryogenic datasets and are different from the results published in \citet{Chase98} and \citet{Bannon05}, because the impacts of the solid phases and of the latent heats are not taken into account in these previous studies, where only the vapour state (prefect gases) of the species are considered.

  \section{\underline{Moist-air specific entropy}.}
     \index{sec25.3}\label{sec25.s_values}

The computation of the specific moist entropy are presented in details in this section, according to the method published in \citet{marquet11}.

Moist air is assumed to be a mixture of four ideal gases at the same temperature\footnote{This hypothesis is not strictly valid in regions where precipitating species are observed at about $T'_w$.
} $T$: dry air, water vapour, liquid-water droplets or ice crystals.
The condensed phases cannot coexist\footnote{They coexist in the real atmosphere and in several GCM, NWP and other models.
However, the mixed phase is not taken into account in this study.
}, except at the triple point ($273.16$~K).
There is no supersaturation and no metastable phase of water\footnote{This 
hypothesis may be easily overcome  by using the affinities defined in \citet{hauf_holler87}, where the relative humidities $e/e_{sw}(T)$ and $e/e_{si}(T)$ may be greater than $1$ in clouds.
} 
(thus, supercooled water is excluded).
If a condensed phase exists, partial pressure of water vapour is equal to the saturation pressure over liquid water ($e_{sw}$)  if $T>0\:^{\circ}$C or ice ($e_{si}$)  if $T<0\:^{\circ}$C.

The specific entropy is written as the weighted sum of the specific partial entropies.
It can be expressed by (\ref{def_s}).
Following \citet{hauf_holler87}, the total-water content $q_t = 1 - q_d = q_v + q_l + q_i$ is used to transform Eq.~(\ref{def_s}) into
\begin{eqnarray}
s  & = & \, q_d \: s_d 
    \: + \: q_t \: s_v 
    \: + \: q_l \: (s_l-s_v)
    \: + \: q_i \: (s_i-s_v) 
    \: .
\label{defS_s_b}
\end{eqnarray}
The differences in partial entropies can be expressed in terms of the differences in specific enthalpy and Gibbs' functions, leading to
\begin{align}
T \: ( s_l-s_v ) & 
  =  \: - \: (h_v - h_l) 
     \: - \: (\mu_l - \mu_v) 
\: , \label{defS_s_cl} \\
T \: ( s_i-s_v ) & 
  =  \: - \: (h_v - h_i) 
     \: - \: (\mu_i - \mu_v) 
\: . \label{defS_s_ci}
\end{align}
When the properties $L_{\mathrm{vap}}= h_v - h_l$, $L_{\mathrm{sub}}= h_v - h_i$, $\mu_l - \mu_v = R_v \: T \: \ln(e_{sw}/e)$ and $\mu_i - \mu_v = R_v \: T \: \ln(e_{si}/e)$ are inserted into Eqs.~(\ref{defS_s_b}) to (\ref{defS_s_ci}), the specific moist-air entropy may be written as
\begin{align}
s  & =   \: q_d \: s_d 
    \: + \: q_t \: s_v 
    \: - \: \left( 
            \frac{L_{\mathrm{vap}}\:q_l 
                + L_{\mathrm{sub}}\:q_i}
                 {T} 
            \right)
  \nonumber \\ 
   & \quad
    \: - \: R_v \left[\: 
              q_l \: \ln\left( \frac{e_{sw}}{e} \right)
            + q_i \: \ln\left( \frac{e_{si}}{e} \right)
              \:\right]
    \: .
\label{defS_s_d}
\end{align}
The second line of Eq.(\ref{defS_s_d}) cancels out for clear-air regions (where $q_l=q_i=0$, but with $q_v$ different from $0$).
It is also equal to zero for totally cloudy air (where $q_l$ and $q_i$ are different from $0$ and $q_v$ is equal to the saturating value), because for liquid-water $q_l \neq 0$ with $e=e_{sw}$ and $\ln(e_{sw}/e)=0$, and for ice $q_i \neq 0$ with $e=e_{si}$ and $\ln(e_{si}/e)=0$.

The first line of Eq.(\ref{defS_s_d}) is then computed with the dry-air and water-vapour entropies expressed as
\begin{align}
s_d  & =   \: (s_d)_r
    \: + \: c_{pd} \: \ln\left( \frac{T}{T_r} \right)
    \: - \: R_d \: \ln\left( \frac{p_d}{(p_d)_r} \right)
 \: ,\label{defS_s_ed} \\
s_v  & =   \: (s_v)_r
    \: + \: c_{pv} \: \ln\left( \frac{T}{T_r} \right)
    \: - \: R_v \: \ln\left( \frac{e}{e_r} \right)
 \: .\label{defS_s_ev}
\end{align}
It is assumed that the specific heats and the gas constants are indeed constant within the atmospheric range of temperature and pressure\footnote{This is valid with a high accuracy for dry air, water vapour and liquid water, but this is not true for ice.
It is however possible to derive a version of moist-air entropy and $\theta_s$ where $c_i(T)$ depends on temperature.
}, with the reference values $(s_d)_r$ and $(s_v)_r$ associated with the references conditions $T_r$ and $p_r=(p_d)_r+e_r$, with $e_r(T_r)$ equal to the saturation pressure at the temperature $T_r$.

When Eqs.~(\ref{defS_s_ed}) and (\ref{defS_s_ev}) are inserted into the first line of Eq.~(\ref{defS_s_d}), and after several rearrangements of the terms which are described in Appendix~B of \citet{marquet11}, the moist-air specific entropy can be written as
\begin{align}
s  & = \: s_r \: + \: {c}_{pd} \: 
               \ln\left( \frac{{\theta}_{s}}
                              {\theta_{sr}} 
                  \right)
    \;  = \: {s}_{\mathrm{ref}} \: + \: {c}_{pd} \: 
               \ln\left( {\theta}_{s} \right)
    . \,
  \label{defS_THs}
\end{align}
The reference potential temperature  ${\theta}_{sr}$ is equal to
\begin{eqnarray}
{\theta}_{sr}   
   & = &
     T_r \;
     \left( \frac{p_0}{p_r}\right)^{\kappa}
     \exp \left( {\Lambda}_r\:q_r \right) \:
     (1+\eta\:r_r)^{\kappa}
  \:  \label{def_THsr}
\end{eqnarray}
and the associated reference entropy ${s}_{\mathrm{ref}}$ is equal to $s_r - c_{pd} \: \ln(\theta_{sr})$, with $s_r = (1-q_r)\:(s_d)_r + q_r\:(s_v)_r$.
The reference specific content and mixing ratio are equal to $q_r=r_r/(1+r_r)$ and $r_r = \varepsilon \: e_r(T_r)/[\:p_r-e_r(T_r)\:]$, where $e_r(T_r)$ is the saturation pressure at $T_r$.
Both $q_r$ and $r_r$  are thus completely determined by the couple ($T_r$, $p_r$).

The moist-air potential temperature ${\theta}_{s}$ appearing in the logarithm of Eq.~(\ref{defS_THs}) is equal to
\begin{eqnarray}
\!\!\!\!\!\!\!\!\!\!\!\!
{\theta}_{s}   \:
   & = & \:
         \: \theta \:
        \exp \left( - \:
                     \frac{L_{\mathrm{vap}}\:q_l + L_{\mathrm{sub}}\:q_i}{{c}_{pd}\:T}
                \right) 
      \; \exp \left( {\Lambda}_r\:q_t \right)
   \nonumber \\
   & &  \times \:
        \left( \frac{T}{T_r}\right)^{{\lambda} \:q_t}
        \left( \frac{p}{p_r}\right)^{-\kappa \:\delta \:q_t}
     \left(
      \frac{r_r}{r_v}
     \right)^{\gamma\:q_t}
      \frac{(1+\eta\:r_v)^{\:\kappa \, (1+\delta \,q_t)}}
           {(1+\eta\:r_r)^{\:\kappa \,\delta \,q_t}}
  \: . \label{def_THs}
\end{eqnarray}
It is defined in terms of the leading order\footnote{This 
first-order $({\theta}_{s})_1$  and higher-order approximations of ${\theta}_{s}$ can be derived mathematically as Taylor expansions of Eq.(\ref{def_THs}) expressed in terms of the small water contents $r_v \approx q_t \ll 1$ (unpublished results). 
This is described with some details at the end of section~\ref{sec25.Thetas_Thetas1}.}
 quantity $({\theta}_{s})_1$ equal to the first line of Eq.(\ref{def_THs}) and equal to
\begin{eqnarray}
 ({\theta}_{s})_1
   & = & \: \theta_l
       \; \exp \left( {\Lambda}_r\:q_t \right)
  \: , \label{def_theta_s1_Betts}
\end{eqnarray}
where 
\begin{eqnarray}
{\Lambda}_r & = & \; \frac{ (s_{v})_r - (s_{d})_r }{ c_{pd}}
  \: \label{def_Lambda}
\end{eqnarray}
is a non-dimensional value depending on the reference entropies of water vapour $(s_{v})_r$ and dry air $(s_{d})_r$.
The liquid-ice potential temperature $\theta_l$ appearing  in Eq.(\ref{def_theta_s1_Betts}) is equal to
\begin{eqnarray} 
  \theta_l
   & = & \: 
        \: \theta \:
        \exp \left( - \:
                     \frac{L_{\mathrm{vap}}\:q_l + L_{\mathrm{sub}}\:q_i}{{c}_{pd}\:T}
                \right)
  \: . \label{def_thetal}
\end{eqnarray}
The quantity $\theta_s$ determined by Eqs~(\ref{def_THs}) to (\ref{def_thetal}) has a rather complex expression, although it is roughly similar to the ones published in \citet{hauf_holler87}, \citet{marquet93} or \citet{emanuel94}.

The reference value ${s}_{\mathrm{ref}} = s_r - c_{pd} \: \ln(\theta_{sr})$ can be computed with $\theta_{sr}$ given by (\ref{def_THsr}) and with the reference values $\Lambda_r$ evaluated  with $(s_v)_r$ and $(s_v)_r$ expressed in terms of the standard values $s^0_v$ and $s^0_d$ computed at $T_0$ and $p_0$, leading to
\begin{align}
(s_d)_r  & =   \: s^0_d
    \: + \: c_{pd} \: \ln\!\left( \frac{T_r}{T_0} \right)
    -  R_d \: \ln\!\left( \frac{(p_d)_r}{p_0} \right)
   = 6777 \: \mbox{J} \: \mbox{K}^{-1} \: \mbox{kg}^{-1}
 \: ,\label{def_sdr} \\
(s_v)_r  & =   \: s^0_v
    \: + \: c_{pv} \: \ln\!\left( \frac{T_r}{T_0} \right)
     -  R_v \: \ln\!\left( \frac{e_r}{p_0} \right)
 = 12673 \: \mbox{J} \: \mbox{K}^{-1} \: \mbox{kg}^{-1}
 \: .\label{def_svr}
\end{align}
The important and expected  result is that most of the terms cancel out and that ${s}_{\mathrm{ref}} = s^0_d - c_{pd} \:\ln(T_0)$.
It can be computed from the Third law value $s^0_d$ given by Eq.(\ref{def_s0d}), leading to ${s}_{\mathrm{ref}} \approx1138.6$~J~K${}^{-1}$~kg${}^{-1}$.
It is thus a constant term and it is independent on the choice of the reference values $T_r$ and $p_r$, provided that $r_r = \varepsilon \: e_r(T_r)/[\:p_r-e_r(T_r)\:]$ is equal to the saturation value at $T_r$.

Since the specific heat for dry-air ${c}_{pd}$ is also a constant term, the moist-air entropy potential temperature ${\theta}_{s}$ given by Eq.~(\ref{defS_THs}) becomes a true equivalent of the moist-air entropy $s$, whatever the changes in dry-air and water species contents may be.
This important property was not observed by the potential temperatures $\theta_S$, $\theta^{\star}$, $\theta_l$ or $\theta_e$ previously defined in \citet{hauf_holler87}, \citet{marquet93} or \citet{emanuel94}, respectively.\footnote{
The notations $\theta_S$ and $\theta^{\star}$ were used in old papers \citet{hauf_holler87}, \citet{marquet93} or \citet{emanuel94} to denote approximate moist-air entropy potential temperatures}

It is possible to write ${\theta}_{s}$ differently, in order to show more clearly that the entropic potential temperature  is independent of the reference values.
The aim is to put together all the terms containing $T_r$, $p_r$ or $r_r$ to form the last line of
\vspace*{-1.5mm}
\begin{align}
{\theta}_{s}   
   & = 
        \: \theta^{\:1+\delta q_t} \;
        \: T^{\:(\lambda-\delta) \:q_t} \;
         \exp \left( - \:
                     \frac{L_{\mathrm{vap}}\:q_l  + L_{\mathrm{sub}}\:q_i}{{c}_{pd}\:T}
                \right)
  \nonumber \\
  & \quad \times \;
      \frac{(1+\eta\:r_v)^{\:\kappa \: (1+\:\delta \:q_t)}}
           {(\eta\:r_v)^{\:\gamma \:q_t}}
    \; \times \;
         \; \exp \left\{ \: \left[ \: {\Lambda}_0\: - \lambda \: \ln(T_0) \: \right] q_t \: \right\} \;
  . \label{def_THs2}
\end{align}
The last exponential term of (\ref{def_THs2}) is computed with $(s_d)_r$  and $(s_v)_r$   evaluated  in Eq.(\ref{def_THs}) in terms of the standard values $s^0_d$ and $s^0_v$ and computed via Eqs.(\ref{def_sdr}) and (\ref{def_svr})  at $T_0$ and $p_0$, leading to
\begin{align}
{\Lambda}_0  \: = \: 
     \left( \: s^0_v -  s^0_d \: \right) 
     / c_{pd} 
   & \approx \: 3.53  
\quad \mbox{and} \quad
{\Lambda}_0 \: - \lambda \: \ln(T_0)  
   \; \approx \:  - \: 1.17  
 \label{def_Lambda_0_T0}
\end{align}
which are clearly independent of $T_r$, $p_r$ or $r_r$.
The standard value ${\Lambda}_0$ corresponds to the definition retained in \citet{hauf_holler87} and is different from the reference one ${\Lambda}_r$ in Eq.~(\ref{def_Lambda}), due to the change of water-vapour partial pressure: $p_0=1000$~hPa in \citet{hauf_holler87}, versus $e_r=6.11$~hPa in ${\Lambda}_r$.

The advantage of Eq.~(\ref{def_THs2}) over Eq.~(\ref{def_THs}) is that all the reference values disappear in Eq.~(\ref{def_THs2}), since they can be replaced by the standard values.
The associated drawback is that all the other terms must be retained in Eq.~(\ref{def_THs2}) for computing accurately the numerical value of ${\theta}_{s}$.

The main difference between the formulation $s=s_{ref}+c_{pd}\:\ln(\theta_s)$ and the other formulations of the moist-air specific entropy is the term ${\Lambda}_r$ defined by Eq.~(\ref{def_Lambda}).
It is a key quantity in the formulations for $\theta_s$.
It is explained in \citet{marquet11} that the advantage of Eq.~(\ref{def_THs}) lies in that it is possible to retain as a relevant approximation of $\theta_s$ the quantity $({\theta}_{s})_1$  defined by  Eq.~(\ref{def_theta_s1_Betts}).
The interesting property is that $({\theta}_{s})_1$ is equal to $\theta_l \:\exp(\Lambda_r\:q_t)$ if $q_i=0$, i.e. a formulation formed by a simple combination of the two Betts' variables $\theta_l $ and $q_t$.

Moreover, if $q_t$ is a constant, ${\theta}_{s}$ varies like the Betts' potential temperature $\theta_l$ up to constant multiplying factors which have no physical meaning when $\ln( {\theta}_{s} )$ is computed in (\ref{defS_THs}).
However, if all the terms in (\ref{defS_THs}) are considered,  if $q_t$ is not a constant and if $q_i$ is different from $0$,  it is expected that the potential temperature $\theta_l$ depending on $q_i$ and the new factor $\exp ( {\Lambda}_r\:q_t )$ depending on the reference entropies may lead to new physical properties.

Differently from $s$ and $\theta_s$ which are independent of the choice of the reference values, both $({\theta}_{s})_1$ and ${\Lambda}_r$  depend on the choice of the couple ($T_r$, $p_r$).
There is no paradox associated with this result.
The fact that numerical values of $({\theta}_{s})_1$ varies with ($T_r$, $p_r$) is compatible with the fact that $\theta_s$  does not depend on them.
The explanation for this result  in that the extra terms in factor of $({\theta}_{s})_1$  in Eq.(\ref{def_THs}) also depend on ($T_r$, $p_r$) and that the change in these terms balance each other in order to give values of $\theta_s$ which are independent of ($T_r$, $p_r$).
The numerical tests  shown in Table~II in \citet{marquet11} validate these results.

   \section{\underline{Moist-air specific thermal enthalpy}.}
     \index{sec25.4}\label{sec25.h_values}

The method used to compute the moist-air specific thermal enthalpy $h$ is described in \citet{marquet15}.
It is similar to the one used in Section~\ref{sec25.s_values} to derive the moist-air specific entropy.

The first step is to write $h$ as the weighted sum (\ref{def_h}).
The properties $L_{\mathrm{vap}} = h_v - h_l$ and $L_{\mathrm{sub}} =  h_v - h_i$, together with $q_d=1-q_t$ and $q_t=q_v+q_l+q_i$, are then used to transform $h$ into the sum
$ h_d +  q_t ( h_v-h_d) -  ( q_l  \: L_{\mathrm{vap}} +  q_i \: L_{\mathrm{sub}} )$.
The next step is to write $h_d = (h_d)_r + c_{pd} \:  (T-T_r) $ and $h_v= (h_v)_r + c_{pv} \: (T-T_r)$ in terms of the reference value $(h_d)_r$ and $(h_v)_r$ expressed at temperature  $T_r$, with $(h_d)_r$ and $(h_v)_r$ to be determined from the standard values $h^0_d$ and $h^0_v$ given by Eqs.(\ref{def_h0d}) and (\ref{def_h0v}).

After rearrangement of the terms the moist enthalpy can be written as
\begin{align}
h \:  & = \: h_{ref}  
\: + \: 
{c}_{pd} \: T
 \: - \:
  L_{\mathrm{vap}}\:q_l  
\: - \:  
  L_{\mathrm{sub}}\:q_i
 \: +\: 
    {c}_{pd}  
\left( \:
  \lambda \; T
 \: + \: 
  T_{\Upsilon} 
 \: \right) \: q_t
  \: . \label{def_h_b}
\end{align}
The term $T_{\Upsilon} = T_r \: [ \: \Upsilon(T_r) - \lambda \: ]$ depends on  ${\Upsilon}(T_r) = [ \: (h_{v})_r - (h_{d})_r \: ]/( c_{pd} \; T_r)$, which is the equivalent of $\Lambda_r$ in the definition of the moist-air entropy potential temperature $\theta_s$. 
It is possible to define an enthalpy temperature $T_h$ by setting $h  = h_{ref} + {c}_{pd} \: T_h$, where ${c}_{pd} \: T_h$ corresponds to 
\begin{align}
T_h \:  & = \: 
 T
 \: - \:
 \left(
 \frac{
  L_{\mathrm{vap}}\:q_l  
  \: + \:  
  L_{\mathrm{sub}}\:q_i
 }{{c}_{pd} }
 \right)
 \: + \: 
\left( \:
  \lambda \; T
 \: + \: 
  T_{\Upsilon} 
 \: \right) \: q_t
  \: . \label{def_Th}
\end{align}

It is shown in \citet{marquet15} that $T_{\Upsilon}$  is independent on $T_r$, even if  ${\Upsilon\!}_r$ depends on $T_r$.
The quantity $h_{ref} $ is also independent of the reference value $T_r$.
Both $h_{ref}$ and $T_{\Upsilon}$ can be expressed in terms of the standard values $h^0_d$ and $h^0_v$ given by Eqs.(\ref{def_h0d}) and (\ref{def_h0v}), leading to the numerical values
\begin{align}
   T_{\Upsilon} \: 
   & = \:
   T_0 \: [ \: {\Upsilon}(T_0) - \lambda \: ] 
       \: = \: 2362  \mbox{~K}
   \: , \label{def_Tr_Upsilon_cste}
\end{align}
and
\begin{align}
   h_{ref}  \: 
   & = \;  
   h^0_{d}  \: - \: {c}_{pd} \: T_0 \; = \: 256  \mbox{~kJ} \mbox{~kg}^{-1}
   \: . \label{def_Href0}
\end{align}

 \section{\underline{Physical properties of the moist-air entropy and ${\theta}_{s}$}.}  
\index{sec25.5}\label{sec25.Physical_properties_S}

     \subsection{Comparisons between ${\theta}_{l}$, ${\theta}_{s}$, $({\theta}_{s})_1$ and ${\theta}_{e}$}
     \index{sec25.5.1}\label{sec25.comparisons_theta_sle}

Let us assume for the moment that $\theta_s$ can be approximated by $({\theta}_{s})_1$.
Comparisons of $({\theta}_{s})_1$ with the well-known potential temperatures $\theta_l$, $\theta_v$ and $\theta_e$ are facilitated by the study of the approximate formulations
\begin{align}
\theta_l  \;  =  \; \theta \: 
    \exp\left(  -   \frac{L_{\mathrm{vap}}\:q_l  }
                             {{c}_{pd}\:T} 
   \right)
 & \approx \; \theta \: 
    \left( 1 + 0 \: q_v - 9 \: q_l \right)
 \: ,\label{def_theta_l} 
\\
\theta_v  \;  = \; \theta \: 
    \left( 1 + \delta \: q_v - q_l \right) \;
  & \approx \;  \theta \: 
    \left( 1 + 0.6 \: q_v - q_l \right)
 \: ,\label{def_theta_v} 
\\
\theta_s \; \approx \; 
(\theta_s)_1  \;  =  \; \theta \: 
    \exp\left( -  \frac{L_{\mathrm{vap}}\:q_l  }
                             {{c}_{pd}\:T} 
             + \Lambda_r \: q_t
     \right)
 & \approx \; \theta \: 
    \left( 1 + 6 \: q_v - 3 \: q_l \right)
 \: ,\label{def_theta_s1}
\\
\theta_e  \;  \approx  \; \theta \: 
    \exp\left( \frac{L_{\mathrm{vap}}\:q_v  }
                             {{c}_{pd}\:T} 
   \right)
 & \approx \; \theta \: 
    \left( 1 + 9 \: q_v \right)
 \: .\label{def_theta_e} 
\end{align}
Only liquid-water is considered in this section, but $q_i$ and $L_{\mathrm{sub}}$ terms may be added in the first three equations if needed.
The approximations are obtained with the properties $\Lambda_r \approx 6$ (see the next section), $\exp(x)\approx (1+x)$ valid for $|x|\ll 1$ and $L_{\mathrm{vap}}/({c}_{pd}\:T) \approx 9$ valid for atmospheric conditions.

\begin{figure}[ht]
\centering
\includegraphics[width=0.36\linewidth,angle=0,clip=true]{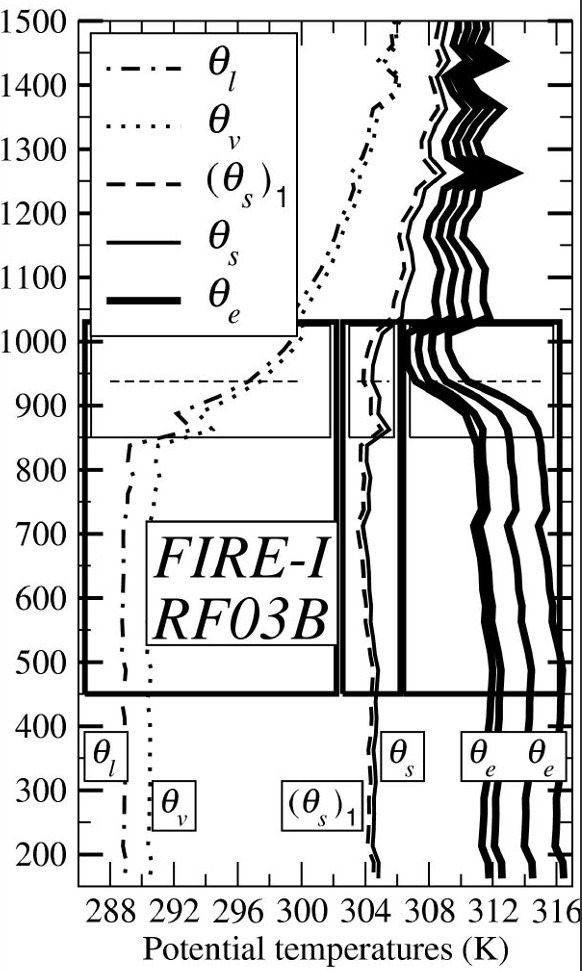}
\includegraphics[width=0.61\linewidth,angle=0,clip=true]{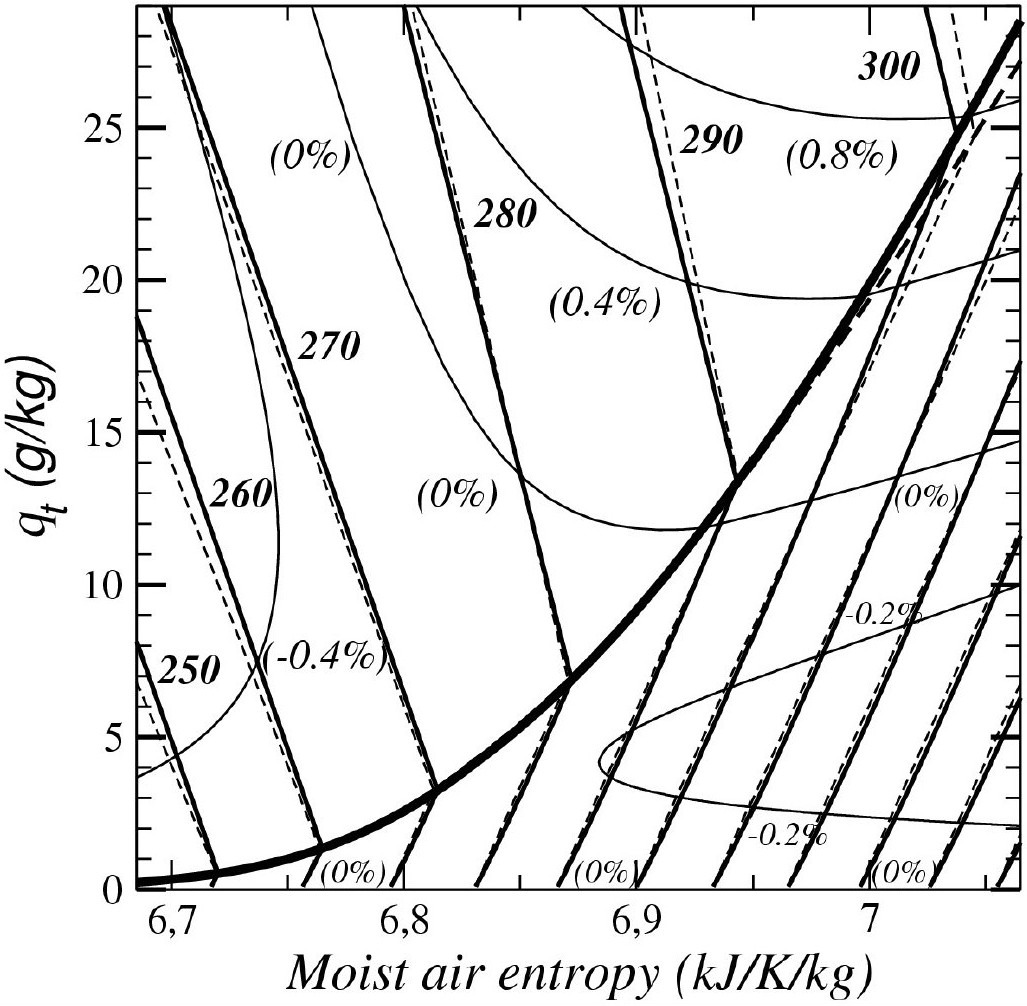}\\
{\hspace{0mm} (a) \hspace{60mm} (b)}
\caption{\it \small 
(a) Vertical profiles of several potential temperature plotted from the FIRE-I (RF03B flight) datasets \citep{marquet11}. 
Different values of $\theta_e$  are plotted by using four different definitions suggested in old papers \citep{Betts73, Bolton80,emanuel94} and a recent one used in ARPEGE-IFS model.
(b) Conservative variable diagram at $900$~hPa, with $q_t$ plotted in ordinates and moist-air entropy computed either with ${\theta}_{s}$ or $({\theta}_{s})_1$ plotted in abscissas.
Isotherms are plotted every $10$~K, with solid lines for ${\theta}_{s}$ and dashed lines for $({\theta}_{s})_1$.
Contours of relative error in moist-air entropy -- i.e. if ${\theta}_{s}$ was replaced by  $({\theta}_{s})_1$ -- are plotted as thin solid lines.}
\label{fig25.5}
\end{figure}

The impact of water vapour ($q_v$)  is zero for $\theta_l$ and very small for $\theta_v$ (factor 0.6).
The factor $9$ is much larger for $\theta_e$, and it is $2/3$ of it for $(\theta_s)_1$.
This means that $\theta_l$ and $\theta_v$ must remain close to $\theta$, whereas $\theta_s$ and $\theta_e$ must differ more from the other potential temperatures.
Moreover, if liquid-water content is discarded ($q_l$ is small in stratocumulus), $(\theta_s)_1$ must be in about a $6/9 = 2/3$ proportion between $\theta_l$ and $\theta_e$.
These consequences are confirmed for the observed dataset depicted in  Fig.(\ref{fig25.5})(a) for the grid-average vertical profiles of the FIRE-I (RF03B case).

    \subsection{Use of $({\theta}_{s})_1$ as an approximation to ${\theta}_{s}$}
    \index{sec25.5.2}\label{sec25.Thetas_Thetas1}

The full expression of ${\theta}_{s}$ given in Eq.(\ref{def_THs}) can be used as such in order to compute ${\theta}_{s}$  numerically, for either under-saturated or saturated moist air.
The interest of searching for a more simple expression for $\theta_s$ is  to better understand the physical meaning of ${\theta}_{s}$ and to facilitate the analytical comparisons with $\theta_l$ or $\theta_e$.

The result shown in \citet{marquet11} is that $({\theta}_{s})_1$ seems to be a relevant approximation for  ${\theta}_{s}$  for $T_r=T_0$ and $p_r=p_0$, with $e_r=e_{sw}(T_0) \approx 6.11$~hPa and $(p_d)_r = p_0 - e_{sw}(T_0)$, differently from the standard conditions where $(p_d)_r = e_r = p_0$.
For these conditions, the numerical value of $\Lambda_r$ is close to $5.87$.
It is possible to appreciate on  Fig.(\ref{fig25.5})(a) how far this expression of $(\theta_s)_1$ (computed with this value $5.87$ for $\Lambda_r$)  is a relevant approximation of $\theta_s$.
This is true from the surface layer up to the level $1500$~m in Fig.(\ref{fig25.5})(a)  and the accuracy of the approximation  ${\theta}_{s} \approx({\theta}_{s})_1$ is even improved above the level $1500$~m, because the total-water content $q_t$ is smaller.
Other comparisons of ${\theta}_{s}$ and $({\theta}_{s})_1$ are available in \citep{marquet11,marquet_geleyn13,marquet14}.

Differences between $(\theta_s)_1$  and $\theta_s$ remains close to $0.4$ to $0.6$~K at all levels for FIRE-I vertical profiles.
The fact that the differences $\theta_s - (\theta_s)_1$ is almost constant with height is an important result if gradients of $\theta_s$ are to be considered (see the Sections~\ref{sec25.moist_BVF} and \ref{sec25.moist_PV} dealing with the moist-air Brunt-V\"ais\"al\"a frequency and  potential vorticity).

Other comparisons between $\theta_s$ and $(\theta_s)_1$ (with $\Lambda_r=5.87$) are shown in the entropy diagram plotted in  Fig.(\ref{fig25.5})(b), where a wide range of saturated and non-saturated conditions are explored at $p=900$~hPa.
The same differences  of the order of $\pm 0.2$~\% or $\pm 0.6$~K are observed for most parts of the diagram.

The same entropy diagram is plotted in \citet{marquet_geleyn13}, but with the set of isotherms computed with the definition of entropy of \citet{Pauluis_Schum_10}, based on the one of \citet{emanuel94} in terms of $\theta_e$.
It is shown that the definition of moist-air entropy in terms of $\theta_s$ or $\theta_e$ leads to incompatible features in the saturated regions, where isotherms corresponds to almost constant values of $\theta_e$ and to clearly decreasing values of $\theta_s$.
These isotherms are also different in the non-saturated region, although they correspond to increasing values of both $\theta_e$ and $\theta_s$.
This demonstrates that the way in which the moist-air entropy is defined may generate important differences in physical interpretation, with $\theta_s$ being the only general formula valid whatever $q_t$ may be.

First- and second-order approximations of $\theta_s$ can be derived by using the Taylor expansion of several terms in the second lines of Eq.(\ref{def_THs}).
The first order expansion of  $(1 + \eta \: r_v)^{[\: \kappa \: (1+\delta \:q_t)\:]}$ for small $r_v \approx q_t$ is equal to $\exp(\: \gamma \: q_t )$.
The term $(r_r/r_v)^{(\gamma\:q_t)}$ is exactly equal to $\exp[\: - \: \gamma \: q_t \: \ln(r_v/r_r) \: ]$.
Other terms depending on temperature and pressure are exactly equal to $\exp[\: \lambda \; q_t \: \ln(T/T_r) \: ]$ and $\exp[\: - \: \kappa \: \delta \: q_t \: \ln(p/p_r) \: ]$.
The last term $(1 + \eta \: r_r)^{(\kappa \: \delta \:q_t)}$ leads to the higher order term $\exp[\: \gamma \:  \delta \: q_t \: r_r \: ] \approx \exp[\: O(q^2_t) \:]$, since $r_r \approx q_t \ll 1$.
The result is that the Taylor expansion of $\theta_s$ can be written as
\begin{eqnarray}
\!\!\!\!\!\!\!\!\!\!\!\!
{\theta}_{s}   \:
   & \approx  & \:
         \: \theta \:
        \exp \left( - \:
                     \frac{L_{\mathrm{vap}}\:q_l + L_{\mathrm{sub}}\:q_i}{{c}_{pd}\:T}
                \right) 
      \; \exp \left( {\Lambda}_{\ast}  \: q_t \right)
   \nonumber \\
   & &  \times \:
      \; \exp \left\{ \: q_t \: \left[ \:  
                \lambda \; \ln\left(\frac{T}{T_r} \right) 
                 \: - \: 
                 \kappa \: \delta \: \ln\left( \frac{p}{p_r} \right) 
                  \:\right]  \:  + \: O(q^2_t) \; \right\}
  \: , \label{def_THs_approx}
\end{eqnarray}
where ${\Lambda}_{\ast} = {\Lambda}_r \: - \: \gamma \; \ln(r_v/r_{\ast})$ and $r_{\ast} = r_r \times \exp(1) \approx 10.4$~g~kg${}^{-1}$.

It is thus possible to interpret $({\theta}_{s})_1$ by the value of ${\theta}_{s}$ obtained with the approximation ${\Lambda}_{\ast} \approx {\Lambda}_r$ and with the second line of Eq.(\ref{def_THs_approx}) neglected.
An improved second order approximation is obtained by taking into account the small term $- \: \gamma \; \ln(r_v/r_{\ast})$ and by using ${\Lambda}_{\ast}$ instead of ${\Lambda}_r$.
The impact of the second line containing terms depending on temperature and pressure leads to a third order correction of $r_{\ast}$ which should be pragmatically increased up to about $12.4$~g~kg${}^{-1}$ for usual atmospheric conditions (unpublished results).

     \subsection{Comparisons with alternative moist-air entropy formulas}
     \index{sec25.5.3}\label{sec25.Pauluis}

The issue whether the Third Law can be applied to atmospheric studies or not, and to determine how this can be managed practically, is an old question.
\citet{Richardson22} (pp.159-160) already wondered if it could be possible to ascribe a value for energy and entropy for a unit mass of (water) substance.
He proposed to take the absolute zero temperature as the zero origin of entropies.
He recognized that the entropy varies as $c_p \:dT/T$ and may make the integral have an infinity where $T=0$~K.
But he mentioned that Nernst had shown that the specific heats tend to zero at $T=0$~K in such a way that the entropy remains finite there.
This is due to Debye's law which is valid for all solids and for which $c_p(T)$ is proportional to $T^3$ (the entropy is thus defined by $ds=c_p\:dT/T$ and is also proportional to $T^3$, a formulation which is not singular for $T$ approaching $0$~K).

However, probably due to lack of available standard values of entropies and energy in the early twentieth century, Richardson did not really use the Third Law.
He suggested to consider the lowest temperature occurring in the atmosphere ($180$~K) as the more practical value for the zero origin of entropies.
This is in contradiction with the above computations based on the Third Law, with the definition depending on $\theta_s$ given by  Eq.~\ref{def_THs} that can be considered as being the moist-air specific entropy imagined by Richardson.

A synthetic view of existing formulations of moist-air entropy derived in \citet{IG73}, \citet{Betts73} and \citet{emanuel94} in terms of $\theta_l$ or $\theta_e$ is suggested in \citet{Pauluis_al_10}.
Two moist-air entropies are defined.
The first one was called  ``moist entropy'' and was denoted by $S_m$.
It is rather written as $S_e$ in (\ref{def_P10_Se}), since it is associated with $\theta_e$.
It can be written as
\begin{align}
  S_e & = \: \left[ \: c_{pd} + q_t \: (c_l-c_{pd} )\: \right] 
         \ln\left( \frac{T}{T_r}\right)
       + \: q_v \:\frac{L_{vap}}{T}
 \nonumber \\
      & \quad
       - \: q_d \: R_d \: \ln\left( \frac{p-e}{p_r-e_r}\right)
       - \: q_v  \: R_v \: \ln\left( \frac{e}{e_{sw}}\right)
   . \label{def_P10_Se}
\end{align}
The second one is given by (\ref{def_P10_Sl}).
It was called  ``dry entropy'' and was denoted by $S_l$ in \citet{Pauluis_al_10}.\footnote{
This is slightly different from the value given in (A3) in \citet{Pauluis_al_10}, because the last term ${e}/{e_{sw}(T)}$ in (\ref{def_P10_Se}) was written as ${e}/{e_{sw}(T_r)}$ in (A3) of \citet{Pauluis_al_10}.
}
It is associated with the potential temperature  $\theta_l$ and is written as
\begin{align}
  S_l & = \: \left[ \: c_{pd} + q_t \: (c_{pv}-c_{pd}) \: \right] 
         \ln\left( \frac{T}{T_r}\right)
       - \: q_l \:\frac{L_{vap}}{T}
 \nonumber \\
      & \quad
       - \: q_d \: R_d \: \ln\left( \frac{p-e}{p_r-e_r}\right)
       - \: q_t  \: R_v \: \ln\left( \frac{e}{e_r}\right)
  . \label{def_P10_Sl}
\end{align}
It is possible to show that the difference between (\ref{def_P10_Sl}) and (\ref{def_P10_Se}) is equal to $S_e - S_l = q_t \: L_{vap}(T_r) / T_r$.\footnote{
This difference is different from (A5) given in \citet{Pauluis_al_10}, because of the change of ${e_{sw}(T)}$ into $e_r = {e_{sw}(T_r)}$ in (A3) of \citet{Pauluis_al_10}.
}
The important feature is that  $S_e - S_l$ depends on the product of $q_t$ by the  constant term $L_{vap}(T_r) / T_r$.
Therefore, since there is only one physical definition for the moist-air entropy, and since $q_t$ is not a constant in the real atmosphere, $S_e$ and $S_l$ cannot represent at the same time the more general form of moist-air entropy.

More precisely, the comparisons between the Third Law based formulation $s(\theta_s)$ given by  Eqs.~ \ref{defS_THs} and  \ref{def_THs}  and the two formulations $S_e(\theta_e)$ or $S_l(\theta_l)$ can be written as
\begin{align}
  s(\theta_s)  & = S_e(\theta_e) + q_t  \left[ \: (s_l)_r - (s_d)_r \: \right] + (s_d)_r
 \:  , \label{def_P10_s_Se} \\
  s(\theta_s)  & = S_l (\theta_l)+ q_t  \left[ \: (s_v)_r - (s_d)_r \: \right] + (s_d)_r
 \:  . \label{def_P10_s_Sl}
\end{align}

The first result is that, if $q_t$ is a constant, then  $S_e$ and $S_l$ can become specialized versions of the moist air entropy associated with the use of the conservative variables $\theta_e$ and $\theta_l$, respectively.
However, even if $S_e$ and $S_l$ are equal to $s(\theta_s)$ up to true constant terms, the constant terms are not equal to zero and they are not the same for $S_e$ and $S_l$.
Moreover, they depend on the value of $q_t$.
Therefore, even if $q_t$ is a constant (for instance for a vertical ascent of a closed parcel of moist air), it is not possible to compare values of $S_e$ or $S_l$ with those for other columns, since the values of $q_t$ and the ``constant'' terms in (\ref{def_P10_s_Se}) and  (\ref{def_P10_s_Sl}) are different from one column to another.

This means that it is not possible to compute relevant spatial average or horizontal gradients of $S_e(\theta_e)$ or $S_l(\theta_l)$, because the link between $\theta_e$ or $\theta_l$ with the moist air entropy $s(\theta_s)$  must change in space.

It is considered in this study that only the Third-Law based formulation $s(\theta_s)$ is general enough to allow relevant computations of spatial average, vertical fluxes or horizontal gradients of moist air entropy.
Indeed,  if $q_t$ is not a constant, then $s = S_e + (s_d)_r$ only if $(s_l)_r = (s_d)_r$.
Similarly, $s = S_l + (s_d)_r$ only if $(s_v)_r = (s_d)_r$.
Therefore, moist air entropy for open system and varying $q_t$ cannot be represented by $S_e(\theta_e)$ or $S_l(\theta_l)$ because the arbitrary choices for the reference entropies $(s_l)_r = (s_d)_r$ or $(s_v)_r = (s_d)_r$ are not compatible with the Third Law of thermodynamic.

It is suggested in \citet{Pauluis_al_10} (Appendix~C) that the weighted sum $S_a = (1-a)\:S_e + a \: S_l $ is a valid definition of the entropy of moist air, where $a$ is an arbitrary constant going from $0$ to $1$.
The adiabatic formulation $S_l$ corresponds to $a=1$ and $\theta_l$, whereas the pseudo-adiabatic formulation $S_e$ corresponds to  $a=0$ and $\theta_e$.
The weighted sum  $S_a$ applied to (\ref{def_P10_s_Se}) and (\ref{def_P10_s_Sl}) leads to
\begin{align}
  s(\theta_s)  & = S_a + (s_d)_r 
   \: + \: q_t \! \left[ \: (s_l)_r - (s_d)_r + a \: \frac{L_{vap}(T_r)}{T_r} \: \right] 
  . \label{def_P10_Sa} 
\end{align}

This result (\ref{def_P10_Sa}) shows that if the term in factor of $q_t$ is not equal to zero, $S_a$ will be different from the Third Law formulation depending on $\theta_s$.
If the value $\Lambda_r \approx 5.87$ derived in \citet{marquet11}  is retained, this factor is equal to zero for $a = [ \: (s_d)_r - (s_l)_r \: ]  \: T_r / L_{vap}(T_r) \approx 0.356$.
This special value represents (and allows the measurement of) the specific entropy of moist air in all circumstances.
No other hypothesis are to be made on the values of the reference entropies, or on adiabatic or pseudo-adiabatic properties, or on constant values for $q_t$.
This provides another explanation for the result observed in sections~\ref{sec25.comparisons_theta_sle}
 and \ref{sec25.Thetas_Thetas1}: the Third-law potential temperature $\theta_s$ is almost in a $(1-a) \approx 2/3$ versus $a\approx 1/3$ position between $\theta_l$ and $\theta_e$. 

The Third Law is not used in \citet{Pauluis_al_10}, where it is mentioned in Appendix~A that ``... the entropy used in atmospheric sciences (...) does not, however, correspond to the absolute entropy based on Nerst's theorem''.
It is explained in this Appendix~A that the entropy of an ideal gas is fundamentally incompatible with Nernst's theorem as it is singular for $T$ approaching $0$~K, due to the term $\ln(T)$.
As explained in \citet{Richardson22}, this statement may be true for ideal gases, but it does not invalidate the application of the Third Law in atmospheric science, since only the more stable solid states and the Debye's law must be considered to apply Nernst's theorem, with finite results at the limit $T=0$~K.

It is mentioned in \citet{Pauluis_al_10} (Appendix~A) that ``A common practice is to (...) set the reference values for the specific entropies of dry air and of either liquid water or water vapour to $0$''.
This is in contradiction with the Third law and the absolute entropies derived in Section~\ref{sec25.Standard_values}.
Really, the Third Law cannot be by-passed for evaluating the general formula of moist air entropy. 
Reference values must be set to the standard ones obtained with zero-entropy for the most stable crystalline form at $T=0$~K.
If $q_t$ is not a constant, the formulations for $s(\theta_s)$, $S_e$, $S_l$ and $S_a$ are thus different and it is claimed that the Third Law formulation $s(\theta_s)$ is the more general one: it is the only formula which can be applied in all circumstances.

  \section{\underline{Applications for the moist-entropy potential temperatures ${\theta}_{s}$} \\
           \underline{and $({\theta}_{s})_1$}.}
    \index{sec25.6}\label{sec25.Applications_thetas}

    \subsection{Isentropic features - Transition from stratocumulus to cumulus}
    \index{sec25.6.1}\label{sec25.Sc_to_Cu}

It is shown in Fig.(\ref{fig25.5})(a) that $(\theta_s)_1$ is  in about a $2/3$ proportion between $\theta_l$ and $\theta_e$.
This confirm the evaluation made in Section~\ref{sec25.comparisons_theta_sle}.
The other important result shown in Fig.(\ref{fig25.5})(a) is that $\theta_s$ is almost a constant from the surface to the very top of the PBL ($1000$~m), including the entrainment region (from $850$ to $1000$~m) where large jumps in $q_t$ and $\theta_l$ are observed.
This means that the whole PBL of this FIRE-I (RF03B) marine stratocumus exhibits a surprising moist-air isentropic state.

\begin{figure}[ht]
\centering
\includegraphics[width=0.8\linewidth,angle=0,clip=true]{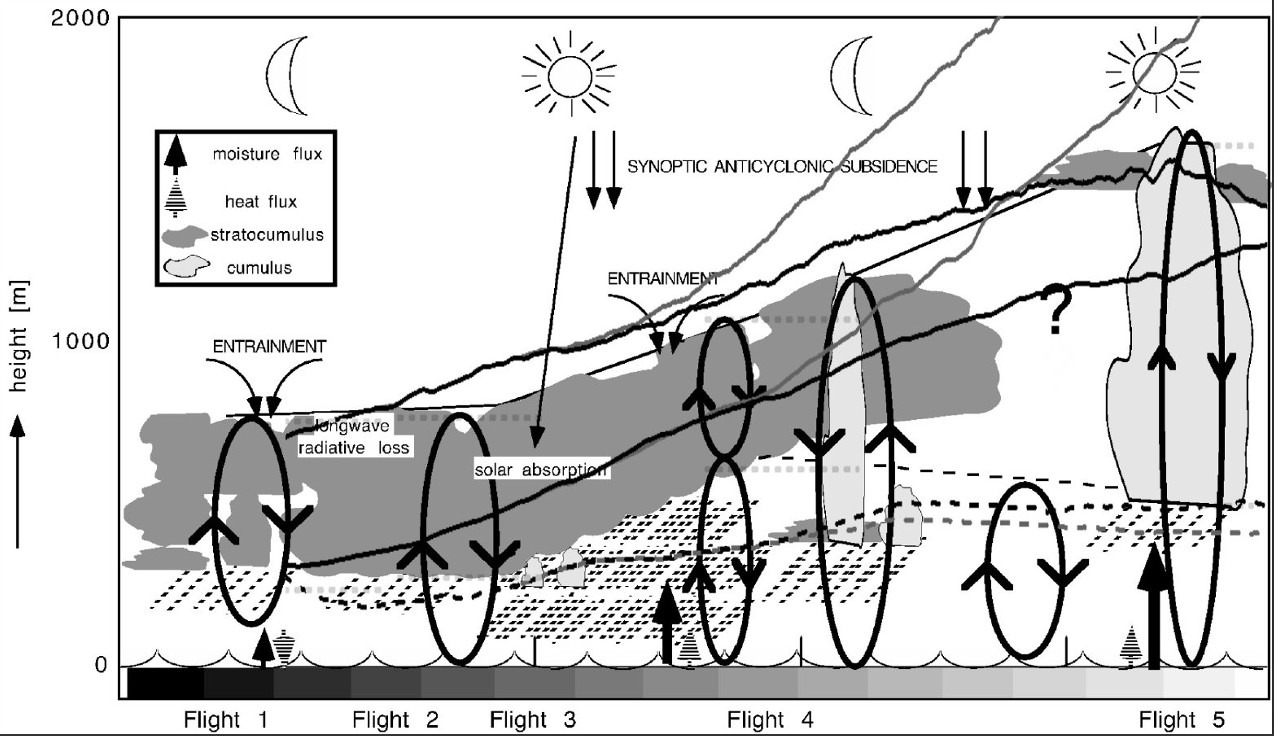}\\
\includegraphics[width=0.99\linewidth,angle=0,clip=true]{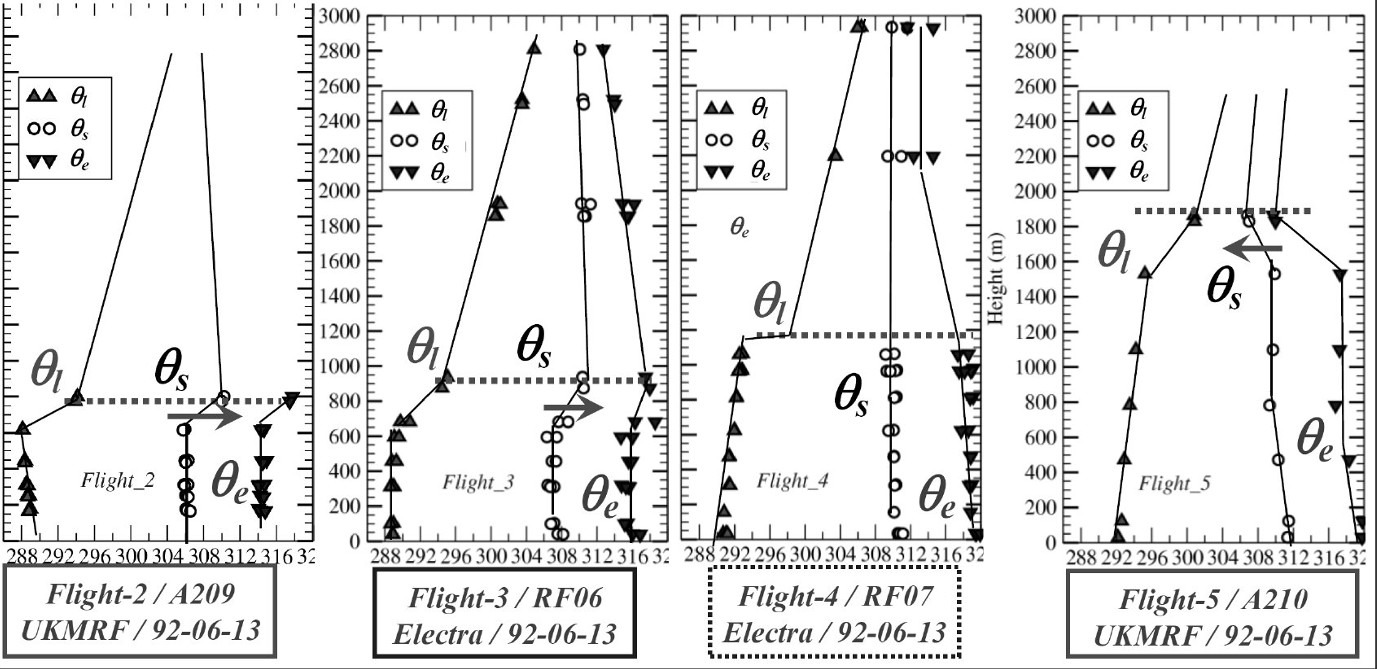}
\caption{\it \small The first ASTEX Lagrangian vertical profiles studied during the EUCLIPSE model intercomparison. Top panel: the schematic of the observed five aircraft flights plotted in   \citep{Roode_Duynkerke_97,Roode_Dussen_2010}. Bottom panel: the vertical profiles of $\theta_l$, $\theta_s$ and $\theta_e$ computed for the flights 2 to 5 (flight 1 is similar to flight 2).}
\label{fig25.6}
\end{figure}

Constant values of the moist-air quantities $s$ and $\theta_s$ are obtained in spite of observed vertical gradients (and top-PBL jumps) in $q_t$ and $\theta_l$, and also in $\theta_e$.
This means that the gradients of the two terms $\theta_l$  and $\exp(\Lambda_r \: q_t)$ combine in such a way that their product remains constant.
This is only observed for that value  $\Lambda\approx 5.87$ determined from the Third Law reference entropies of dry air and water vapour.
It is not valid for the value $\Lambda_r = 0$ (leading to $\theta_l$) or the value $\Lambda_r = L_{\mathrm{vap}} / (c_{pd}\:T) \approx 9$ (leading to $\theta_e$).

It is shown in \citet{marquet11} that the moist  isentropic  feature (constant values of  $\theta_s$) is also observed for other FIRE-I profiles (radial flights 02B, 04B and 08B) and for the vertical profiles of other marine stratocumulus (ASTEX, EPIC, DYCOMS).
This is the confirmation that the Third Law based moist-air entropy defined with $\theta_s$ can reveal unexpected new and important physical properties which have not been  revealed by the previous studies based on $\theta_l$ or $\theta_e$.
Since the moist air isentropic feature is observed in the marine stratocumulus, there is a hope that this very special  property might be taken into account for building -- or improving -- moist turbulent parameterizations, and possibly shallow convection ones (both acting in the cloudy parts of these marine stratocumulus), including the top-PBL entrainment processes.

In order to determine if these isentropic patterns are also observed during the transition from stratocumulus to cumulus (or  how they are modified), several vertical profiles of $\theta_l$,  $\theta_s$ and $\theta_e$ are plotted in  Fig.(\ref{fig25.6}) for the aircraft measurements realized during the first ASTEX Lagrangian and retained in the EUCLIPSE model intercomparison study \citep{Roode_Duynkerke_97,Roode_Dussen_2010}.
The moist-air isentropic feature (constant values of $\theta_s$) is indeed observed for the Flight~4, with an increase of moist-air entropy at the top-PBL of  flights~2 and 3, and a decrease in  $\theta_s$ for the (cumulus) flight~5.

The pattern of $\theta_l$ are almost the same for all profiles, with positive top-PBL jumps that cannot be differentiated at first sight.
The pattern of $\theta_e$ are more reactive than those of $\theta_s$, with a ``neutral'' feature in between flights~3 and 4.
For this aspect, too, $\theta_s$ seems to be in a intermediate position between $\theta_l$ and $\theta_e$.
The sketch of cumulus profile depicted in Flight~5 is compatible with the other cumulus profiles of $\theta_l$, $\theta_s$ and $\theta_e$  (unpublished results for ATEX, GATE, BOMEX and SCMS).

It is explained in next section that a neutral vertical gradient of $\theta_s$ seems to correspond to the $\Delta_2$ CTEI criterion line.
This may demonstrate a new possibility to analyze the strength of a top-PBL inversion: it is (stable / neutral / unstable) if (positive / no / negative)  top-PBL jump in $\theta_s$ are observed, respectively.
Here is a new interesting property for the moist-air entropy expressed in terms of $\theta_s$, since it allows a rapid and graphical analysis of entrainment instabilities, and thus an estimation of evolution of clouds.

    \subsection{Top-PBL entrainment and CTEI criterion}
    \index{sec25.6.2}\label{sec25.CTEI}

The concept of cloud-top entrainment instability (CTEI) is defined in \citet{Lilly68}, \citet{Randall80} and \citet{Deardorff80} in order to find an explanation for the stratocumulus cloud breakup.
It is suggested that critical values exist for the top PBL jumps in $\theta_e$ or $\theta_l$,  hereafter denoted by $(\Delta \theta_e)_{\mathrm{crit}}$ and $(\Delta \theta_l)_{\mathrm{crit}}$, values above which a mixed air parcel will be positively buoyant with respect to the cloud top entrainment processes. 

Instabilities are thus expected if $\Delta \theta_e$ is smaller than the value  $(\Delta \theta_e)_{\mathrm{crit}} = k_{RD} \: L_{\mathrm{vap}}\: \Delta q_t  \: / c_{pd} < 0$, with $\Delta q_t$ equal to the top-PBL jump in $q_t$.
The  Randall-Deardorff parameter $k_{RD}$  has been evaluated to $0.23$ in the first studies and it varies between $0.18$ and $0.7$ in the literature.
It is possible to express the CTEI criterion in the conservative diagram ($\theta_l$, $q_t$), leading to the  buoyancy reversal criterion line usually denoted by $\Delta_2$ \citep{RW07}.
The corresponding threshold is equal to $(\Delta \theta_l)_{\mathrm{crit}} = - \: (1/k_{L}) \: L_{\mathrm{vap}}\: \Delta q_t  \: / c_{pd} > 0$, where $k_{L}\approx1/(1-k_{RD})$  is the Lilly parameter  \citep{Lilly68}.

The fact that moist-air entropy and $\theta_s \approx (\theta_s)_1$ are almost conserved within the whole PBL of marine stratocumulus may help to give a new insight for the CTEI criterion.
Indeed, the property $\Delta(\theta_s)_1 = 0$ corresponds to the value $k_{RD} = 1 - c_{pd} \: \Lambda_r \: \theta / L_{\mathrm{vap}} \approx 0.29$ for $\Lambda_r=5.87$ \citep{marquet11}.
It seems that higher values up to $k_{RD}=0.35$ are associated to the use of the exact formulation $\theta_s$ to compute $\Delta(\theta_s) = 0$ (unpublished results).
These Third-Law based values of $k_{RD}$ varying between $0.29$ and $0.35$ are close to the one derived by Deardorff and Randall ($0.23$) and are located within the range of the published ones ($0.18$ to $0.7$).

It can be inferred that the CTEI mechanism seems to be controlled by a neutral (zero) top-PBL jump in moist-air entropy (in fact by an isentropic state of the whole PBL) without additional assumption needed to derive this value of the Randall-Deardorff parameter $k_{RD}$.
The potential temperatures $\theta_l$ or $\theta_e$ cannot represent changes in moist-air entropy or isentropic processes for marine stratocumulus, which are open systems where $q_t$ is rapidly varying with height.
Only $\theta_s$ can reveal isentropic patterns for such moist-air open systems.

     \subsection{Moist-air turbulence parameterization}
     \index{sec25.6.3}\label{sec25.moist_turb}

Moist PBL of marine stratocumulus are considered as paradigm of moist turbulence.
The property that moist-air entropy (and $\theta_s$) is almost uniform in the whole PBL of marine stratocumulus, including the top-PBL region, is thus an important new result.

The first possible consequence is that it could be meaningful to replace the pair of Betts' variables ($\theta_l$, $q_t$) by the pair ($\theta_s$, $q_t$).
If $\theta_s$ is approximated by $({\theta}_{s})_1 = \theta_l \: \exp(\Lambda_r \:q_t)$, the vertical turbulent fluxes of $(\theta_s)_1$, $\theta_l$ and $q_t$ are then approximated by $\overline{w' \theta'_s} \approx \overline{w' (\theta'_s)_1} = \exp(\Lambda_r \:q_t) \: \overline{w' \theta'_l} + (\theta_{s} \:\Lambda_r) \: \overline{w' q'_t}$.

If linear flux-gradient relationships $\overline{w' \psi'} = - K_{\psi} \: \partial \overline{\psi} / \partial z$ are assumed to be valid for $\psi = (\theta_s)_1$, $\theta_l$ or $q_t$, they define  corresponding exchange coefficients $K_s$, $K_h$ and $K_q$.
It is then worthwhile to compare the two methods: either by computing $\overline{w' (\theta'_s)_1}$ by using $- K_{s} \: \partial \overline{(\theta'_s)_1} / \partial z$, or by using the above weighted sum of the vertical flux $\overline{w' \theta'_l}$ and $\overline{w' q'_t}$, with these fluxes expressed with the corresponding flux-gradient relationships.

The two methods are only equivalent if $K_s = K_h$ and $K_s = K_q$, this implying $K_h  = K_q$.
This equality is often assumed to be valid in numerical modelling, where Lewis number $L_e = K_h / K_q \approx 1$. 
These results are however based on old papers \citep{Swinbank_Dyer_67,Webb_70,Dyer_Hicks_70,Oke_70} and they are sometimes invalidated in other old articles \citep{Blad_Rosenberg_74,Warhaft_76,Verma_al_78,Brost_79}.
More recent studies based on LES outputs suggest that  $K_h$ could be different from $K_q$ \citep{Stevens_al_2001,Siebesma_al_2003}.
Moreover, neutral bulk coefficients are different for heat and water in some surface flux parameterization over oceans, like in ECUME/SURFEX \citep{Belamari_2005,Belamari_Pirani_2007,Masson_SURFEX_2013,LeMoigne_ECUME_2013}.

The two methods are different if either $K_s \neq K_h$ or $K_s \neq K_q$, in which case it is not equivalent to use the pair of Betts' variables or the pair ($\theta_s$, $q_t$).
This could provide a clear possible interest to express the turbulent scheme with the moist-air entropy, since $\theta_l$ is only an approximation of $(\theta_s)_1$ for the case of constant values of $q_t$.

Other possible applications for turbulent parameterizations and shallow convection may be based on the use of moist-air Richardson numbers, equal to the moist-air square of the Brunt-V\"ais\"al\"a Frequency divided by the squared wind shear $R_i = N^2/S^2$.
It is in this context that the results about moist-air definition of $N^2(C)$ recalled in Section~{\ref{sec25.moist_BVF}} may provide new insights for computations of $R_i(C)$.
The transition parameter $C$ might be somehow generalized to grid-cell concept and to sub-grid variability, with $C$ considered as the proportion of an air parcel being in saturated conditions.
This could lead to extended grid-cell interpretations for the local formulation of $N^2(C)$ given by Eq.(\ref{def_N2C}).
One could also imagine possible bridges with the programme followed by \citet{Lewellen2004}, where a parameter $\Hat{R}$  is introduced to mix the ``dry and wet limits'' in the context of computation of moist-air buoyancy fluxes.

     \subsection{The  moist-air squared Brunt-V\"ais\"al\"a Frequency}
     \index{sec25.6.4}\label{sec25.moist_BVF}

One of the applications of the moist-air entropy defined in terms of $\theta_s$ given by Eq.(\ref{def_THs}) is to compute the moist-air adiabatic lapse rate and squared Brunt-V\"ais\"al\"a Frequency (BVF).
Indeed, they both depend on the definition of reversible and adiabatic displacements of a parcel of air, that is to say at constant value of entropy.
The squared BVF can be defined for the dry air by $N^2 = (g/\theta)\: \partial \theta / \partial z$, or equivalently in terms of the dry-air entropy by $N^2 = (g/c_{pd})\: \partial s / \partial z$, since $s = c_{pd} \ln(\theta)$ up to a constant term.
Therefore, the  way the entropy is defined may modify the formulation of both the adiabatic lapse rate and the squared BVF.
This is the initial motivation for using the formulation $\theta_s$ for searching a moist-air generalization for the dry-air expression.

Moist-air versions have been derived by using different methods \citep{LsEi74,DK1982,emanuel94}.
Unsaturated and saturated moist-air new versions of $N^2$ have been recently revisited in \citet{marquet_geleyn13}, where the moist-air entropy is defined in terms of the more general formulation for $\theta_s$ given by Eq.(\ref{def_THs}).
Additional results are described in \citet{marquet_geleyn13} concerning the computation of  the adiabatic lapse rates.
A synthetic view of these results is presented in this Section.

The adiabatic lapse rates are defined in \citet{marquet_geleyn13} in terms of a coefficient $C$ which labels the unsaturated  case by $C=0$ and the saturated case by $C=1$.
This transition parameter allows compact writings valid for both unsaturated and saturated moist air.
It is worthwhile to notice that the unsaturated moist-air is different from the dry-air because it includes  water vapour (with relative humidity lower than $100$\%).

The adiabatic lapse rate computed with $\theta_s$ can be written as
\begin{align}
   \Gamma(C)\: 
   & = \:
 \frac{g}{c_p} \: \times \: M(C)
   \: , \label{def_Lapse_rate}
\end{align}
where
\begin{align}
 M(C) & = \:
 \frac{1+D_C}{1+F(C)\:D_C}
   \: , \;\;\;
   D_C \: = \:
 \frac{L_{\mathrm{vap}}(T) \: r_{sw}}{R_d\:T}
   \: , \label{def_Lapse_rate_D}\\
   F(C)\: 
   & = \:
 1 \: + \:
 C \:
 \left[
 \frac{L_{\mathrm{vap}}(T)}{c_p\:T} \:
 \frac{R}{R_v}
 \: - \: 1
 \right]
   \: . \label{def_Lapse_rate_F} 
\end{align}
Since $F(0)=1$ and $M(0)=1$, the unsaturated adiabatic lapse rate is exactly equal to $\Gamma(0)=g/c_p$, where $c_p$ is the moist value depending on the dry-air and water-vapour contents.

The saturated adiabatic lapse rate is slightly different from those derived in \citet{DK1982} and \citet{emanuel94}.
It is equal to $\Gamma(1)$,  leading to
\begin{equation}
   \Gamma_{sw}
   \; = \;  \frac{g}{c_p} \: \times \: 
   \left[ \:
   \:  \dfrac{1+ \dfrac{L_{\mathrm{vap}}(T) \: r_{sw}}{R_d\:T}}
                 {1+ \dfrac{R\: L^2_{\mathrm{vap}}(T) \: r_{sw}}{R_d\:c_p\: R_v \: T^2}}
  \: \right]
   \: . \label{def_Gamma_sw}
\end{equation}
\vspace{0.5mm}

This formulation enables comparison with previous formulations derived in  \citet{DK1982} and \citet{emanuel94}.
The numerator $1+({L_{\mathrm{vap}} \: r_{sw}})/({R_d\:T})$ is multiplied by a term $(1+r_t)$ in previous studies, where the denominators are also different.
The term $g/c_p$ is replaced by $g/c_{pd}$ in \citet{DK1982} and by $g/c^{*}_p$  in \citet{emanuel94}, where $c^{*}_p=c_{pd}+c_{pv}\:r_{sw}$.

The squared BVF is defined in  \citet{marquet_geleyn13} by the general formula
\begin{align}
  N^2  & = \: - \: \frac{g}{\rho} \:
                \left( \;
                  \left.
                  \frac{\partial \rho}{\partial s}
                  \right|_{p,q_t} \:
                \frac{\partial  s  }{\partial z  }
                \:\; + \:\;
                  \left.
                  \frac{\partial \rho}{\partial q_t}
                  \right|_{p,s} \:
                \frac{\partial q_t }{\partial z  }
                \; \right)
  \: , \label{def_new_N2_moist}
\end{align}
in terms of the vertical gradients of moist-air entropy and water content.
The terms in factor of the gradients are obtained after long computations, because $s$ is defined in terms of $\theta_s$ given by Eq.(\ref{def_THs}).
The result can be written by using the same definitions for $\Gamma(C)$, $M(C)$, $D$ and $F(C)$ described in Eqs.(\ref{def_Lapse_rate}) to (\ref{def_Lapse_rate_F}).
The same transition parameter $C$ is used to express with a single formula the unsaturated ($C=0$) and the saturated ($C=1$) cases.
\begin{align}
   N^2(C)
   & = \: 
   \Gamma(C) \; \frac{\partial s}{\partial z}
   \: + \: g \: \frac{\partial \ln(q_d)}{\partial z}
  \nonumber \\
  & + \:
  \Gamma(C) \; 
   \left[\:
    c_p \: (1+r_v) \:  \frac{R_v}{R} \: F(C) 
   \: - \:
   c_{pd} \: ( \Lambda_r + \Lambda_v )
   \:\right]
   \frac{\partial q_t }{\partial z}
   \: , \label{def_N2C}
\end{align}
where $\Lambda_v$ is an additional term defined in \citet{marquet_geleyn13}.
The new impact of $F(C)$ in the term $\; c_p \: (1+r_v) \:  ({R_v}/{R}) \: F(C) \:$ in the second line is to replace the unsaturated value $(1+r_v)\: c_p \: R_v / R$ if $C=0$ by the saturated version $(1+r_{sw}) \: L_{\mathrm{vap}}\:/\: T$ for $C=1$.

The unsaturated version ($C=0$) leads to the result
\begin{align}
   N^2_{ns}
   & = \: 
   \frac{g}{\theta_v}
   \: \left[ \:
         \frac{\partial \theta_v}{\partial z}
        \: - \:
         \frac{g}{c_p} \: \frac{\theta}{T} \: (\lambda - \delta) \: q_v
     \: \right]
   \: . \label{def_N2_ns}
\end{align}
This expression is close to the expected result $(g/\theta_v) \: ({\partial \theta_v}/{\partial z})$, because the counter-gradient term is much smaller (typically one hundredth of the other).
The saturated version is more complicated, but it is compared in \citet{marquet_geleyn13} with the older results derived in \citet{DK1982} and \citet{emanuel94}, where  $N^2$ was based on a moist-air entropy mainly expressed in terms of $\theta_e$, instead of $\theta_s$.

The important feature is not the numerical value of $N^2$.
The important property is the way how $N^2$  is separated in Eq.(\ref{def_N2C}) into the three terms depending on vertical gradients of $s$, $q_d$ and $q_t$.
This separation depends on the way the moist-air entropy is defined.

Following \citet{PH02a,PH02b}, the first term depending on $s$ can be interpreted as a conversion term between the turbulent kinetic energy and the available potential energy.
The second term depending on $q_d$ is well-known and was already included in \citet{LsEi74} and in other studies. 
It corresponds to the ``total water lifting effect'' (conversion between the turbulent kinetic energy and the potential energy).  
The third term depending on $q_t$ is different from those derived in previous studies, simply because the moist-air entropy is defined differently.
It corresponds to new $\Lambda_r$-scaled differential expansion and latent heat effects, associated with the use of $\theta_s$ instead of $\theta_e$ or $\theta_l$.

     \subsection{The moist-air potential vorticity PV$_s$}
     \index{sec25.6.5}\label{sec25.moist_PV}

Another logical application of the moist-air entropy defined in terms of $\theta_s$ given by Eq.(\ref{def_THs}) is the computation of the moist-air potential vorticity (PV).

The choice of the conservative variable to enter the definition of PV was open in the seminal papers of \citet{Ertel1942} -- in English in \citet{Schubert_al_2004} -- and \citet{HMR85}.
It was mentioned that the modern choice of any of the potential temperature $\theta$ \citep{Ertel1942,HMR85}, $\theta_w$ \citep{BH79}, $\theta_v$ \citep{Schubert_al_2001}, $\theta_{es}$ or $\theta_e$ \citep{emanuel94} might be replaced by the entropy variable like the dry-air value $c_{pd} \: \ln(\theta) + \mbox{cste}$.
It was thus tempting to study the properties of the moist-air PV computed starting with $c_{pd} \: \ln(\theta_s) + \mbox{cste}$.
The results are published in \citet{marquet14}

The description of the meteorological properties of PV and the possibility of associated invertibility principle, or not, are beyond the scope of this book, which is focussed on convective processes.
However, those regions where negative values of PV are associated with positive gradients of the associated potential temperature correspond to criterias for possible slantwise convection instabilities \citep{BH79}.
It is shown in \citet{marquet14} that these criteria are more clearly observed with $\theta_s$ than with $\theta_{es}$ or $\theta_e$.
This result may encourage more systematic studies of slantwise convection with the help of the moist-air potential vorticity denoted by PV$_s$ and based on PV$(s)$.

  \section{\underline{Conservation properties of ${\theta}_{s}$}.}
    \index{sec25.7}\label{sec25.Conservative_theta_s}

    \subsection{Conservative properties of ${\theta}_{s}$ for pseudo-adiabatic processes}
    \index{sec25.7.1}\label{sec25.Thetas_Pseudo_adiab}

It is easy to use $\theta_s$ to determine whether or not a process is isentropic, i.e. by analyzing whether or not  $\theta_s = Cste$.
It is however important to explain why the moist-air entropy increases if pseudo-adiabatic processes are involved, and to compute precisely the change in $\theta_s$ associated with either reversible and adiabatic, or irreversible and pseudo-adiabatic, processes.

Let us first consider pseudo-adiabatic processes for open systems associated with an ascending parcel of just-saturated moist air of mass $m=m_d+m_v$, with no liquid water retained within the parcel.
The pseudo-adiabatic processes can be understood as a series of three elementary steps.

The first step is an infinitesimal and isentropic rising of this parcel, with the appearance of liquid water by condensation processes.
The new mass is equal to $m=m_d+(m_v+dm_v)+dm_l$, with $dm_v=-dm_l$ by virtue of the conservation of the mass.
During the first step the moist-air entropy is assumed to be a constant.
The initial entropy is equal to $S_i = m_d \: s_d + m_v \: s_v + 0 \: \times \:s_l$, whereas the final entropy is equal to $S_f = (m_d+dm_d) \: (s_d+ds_d) + (m_v+dm_v) \: (s_v+ds_v) + dm_l\:(s_l+ds_l)$.
If the second order terms are discarded, the change in entropy divided by $m_d$ is then equal to
\begin{equation}
  (S_f - S_i) / q_d \: = \: 0 \; = \; ds_d + r_{sw}\:ds_v+ (s_v-s_l)\:dr_{sw}
  \: . \label{def_equa_pseud_adiab}
\end{equation}
The water-vapour mixing ratio is equal to its saturation value $r_{sw}=m_v/m_d$ in pseudo-adiabatic processes.
The water-vapour and dry-air specific entropies $s_v$ and $s_d$ must be evaluated at the temperature $T$ and at the saturated partial pressures $e_{sw}(T)$ and $p-e_{sw}(T)$, respectively.

The second step is a removal of the condensed liquid water $dm_l=-dm_v$ by falling rain, with a total mass  reduced to $m_d+m_v-dm_l$ and with an amount of entropy $dm_l\: s_l$ withdrawn by the precipitations.
The consequence of this second step is that $r_t$ is equal to $r_{sw}$ during the whole ascent.
The paradox is that this step does not mean that the specific entropy will decrease.
It is shown hereafter that $s$ increases during pseudo-adiabatic ascent.

The third step is to consider a  ``resizing'' of the parcel of mass $m_d+m_v-dm_v$ into a bigger volume of mass equal to its initial value $m=m_d+m_v$.
The aim of this step would be to compute specific quantities, i.e. for a constant unit mass of moist air (not of dry-air).

The change in specific entropy ($s$) associated with the pseudo-adiabatic processes can be computed by rewriting  Eqs.~(\ref{def_equa_pseud_adiab}) as $0 = d(s_d+r_{sw}\:s_v) -s_l\:dr_{sw} $.
The term into parenthesis is equal to the just-saturated moist-air entropy expressed per unit of dry air ($s/q_d$).
The result is that pseudo-adiabatic processes correspond to $d(s/q_d) = s_l\:dr_{sw}$.
Since the property $1/q_d = 1/(1-q_{sw})=1+r_{sw}$ is valid for just-saturated moist air, it can be used to derive the differential equation
\begin{equation}
ds 
\; = \;
   c_{pd} \: \frac{d\,\theta_s}{\theta_s}
\; = \; 
   ( s - s_l ) \: \left[ \: \frac{- \: dr_{sw}(T,p)}{1+r_{sw}(T,p)} \: \right]
  \: . \label{def_s_pseud_adiab}
\end{equation}
\vspace*{1mm}
It can be deduced from Eq.~(\ref{def_s_pseud_adiab}) that the change in specific moist-air entropy is positive for upward pseudo-adiabatic displacements, since they are associated with the properties $dr_{sw}(T,p)<0$ and $s > s_l$.
The first property is demonstrated for $T$ decreasing with height within the updraft.
The second property is displayed because $q_d \gg q_{sw}$ and thus $s=q_d\: s_d + q_{sw}\:s_v$ is close to the dry-air entropy $s_d$.
Moreover, according to the curves depicted in Fig.~\ref{fig25.3}, $s_d$ is always greater than $s_l$ for the atmospheric range of temperatures.
It is worthwhile noticing that the impact of change in pressure from $p_0$ to $p_d$ in $s_d$ does not modify so much the values of $s_d(T,p_d)$, leading to values of $s_l(T)$ which are always smaller than those of $s$.

The physical meaning of Eq.~(\ref{def_s_pseud_adiab}) is that the impact of the precipitations is to remove liquid-water entropy during pseudo-adiabatic processes, to be replaced by the local value of the moist-air entropy. 
This last action can be interpreted as a detrainment process.
It corresponds to the third step described above, leading to a ``resizing'' of the moist-air parcel in order to be able to compute specific quantities expressed ``per unit of moist air'', with the need to replace the lost mass of liquid water by an equal mass of moist air.
For these reasons, moist-air entropy and $\theta_s$ must be increasing with height in regions where pseudo-adiabatic conditions prevail.

It can be shown that Eq.~(\ref{def_equa_pseud_adiab}) exactly corresponds to the differential equation 
\begin{equation}
 \left( c_{pd} + r_{sw} \: c_l \right) 
 \frac{dT}{T}
 \; - \;
 R_d \: \frac{dp_d}{p_d}
 \; + \;
 d\left( \frac{r_{sw}\:L_{\mathrm{vap}}}{T} \right) 
  \; = \; 0
  \: . \label{def_equa_theta_pw}
\end{equation}
This differential equation is the same as those derived in \citet{Saunders57} or in the \citet{Smithsonian66} Tables to define the water-saturation pseudio-adiabats.
Equation~(\ref{def_equa_theta_pw}) is derived from Eq.~(\ref{def_equa_pseud_adiab}) with the use of the Kirchhoff and Clausius-Clapeyron equations recalled in section~\ref{sec25.Moist_air_thermo}, together with the equalities $ds_d=c_{pd}\:dT/T - R_d \:dp_d/p_d$, $ds_v=c_{pv}\:dT/T - R_v \:de_{sw}/e_{sw}$ and $s_v-s_l=L_{\mathrm{vap}}/T$.
The integration of Eq.(\ref{def_equa_theta_pw}) leads to the definition of the wet-bulb pseudo-adiabatic potential temperature ${\theta}'_{w}$ and the equivalent version ${\theta}'_{e}$.

This is a confirmation that the Third-Law definition of $\theta_s$ does not modify the definition of ${\theta}'_{w}$ (nor ${\theta}'_{e}$), since Eq.(\ref{def_equa_theta_pw}) can be derived from Eq.(\ref{def_equa_pseud_adiab}) and without the use of Eqs.(\ref{defS_THs})-(\ref{def_thetal}) which define $s(\theta_s)$.
The two kinds of potential temperatures correspond to different properties: $\theta_s$ always represents the moist-air entropy (for open or closed systems, for adiabatic or diabatic processes), whereas ${\theta}'_{w}$ and ${\theta}'_{e}$ are conservative properties for pseudo-adiabatic processes only.

    \subsection{Conservative properties of ${\theta}_{s}$ for adiabatic or isentropic processes}
    \index{sec25.7.2}\label{sec25.Thetas_adiab}

Let us consider the adiabatic conservation laws.
The equation defining the water-saturation reversible adiabats are written in \citet{Saunders57} and \citet{Betts73} as
\begin{equation}
 d\left(\frac{s}{q_d}\right)
  \: = \:
 \left( c_{pd} + r_t \: c_l \right) 
 \frac{dT}{T}
 \; - \;
 R_d \: \frac{dp_d}{p_d}
 \; + \;
 d\left( \frac{r_{sw}\:L_{\mathrm{vap}}}{T} \right)
  \: = \: 0
  \: . \label{def_satur_adiabat}
\end{equation}
In comparison with (\ref{def_equa_theta_pw}), $c_{pd} + r_{sw} \: c_l$ is replaced by $c_{pd} + r_t \: c_l$ in (\ref{def_satur_adiabat}).
It is assumed in \citet{Betts73} that Eq.~(\ref{def_satur_adiabat}) corresponds to the conservation of the entropy per unit of dry air, namely $d(s/q_d)=0$, with the hypotheses that $q_d$ and $q_t=1-q_d$ are constant and that $q_i=0$.

The important feature is that the same Eq.(\ref{def_satur_adiabat}) can be obtained in a more straightforward way, by differentiating $s/q_d$ and with $s$ given by Eq.(\ref{defS_s_b}), providing that  the same assumptions of constant values for $q_d$ and $q_t=1-q_d$ and of $q_i=0$ prevail.
Therefore, the water-saturation reversible adiabats can be defined by constant values of the Third-law-based specific entropy $s(\theta_s)$, or equivalently by  $\theta_s=Cste$.

However, the moist-air isentropic equation $ds=c_{pd}\:d\theta_s/\theta_s=0$ is more general than Eq.~(\ref{def_satur_adiabat}), since it can be derived  analytically, without the assumption of constant $q_t$ and $q_d=1-q_t$ and is thus valid for open systems and for varying values of $q_t$.
Therefore, the Third-Law potential temperature $\theta_s$ and the corresponding moist-air entropy $s$ defined by Eqs.(\ref{def_THs}) and (\ref{defS_THs}) can  be interpreted as the general integral of $ds=0$.

It is the quantity $q_d$ times Eq.~(\ref{def_satur_adiabat}) that is integrated via some approximations in \citet{Betts73} and \citet{BettsDugan73}, to arrive at the definitions of the well-known liquid-water and saturation equivalent potential temperature ${\theta}_l$ and ${\theta}_{es}$, respectively.
The water-saturation reversible adiabatic conservative properties verified by ${\theta}_s$, ${\theta}_l$ and ${\theta}_{es}$ correspond to the formal equations 
\begin{eqnarray}
  d s
 \: = \: 0  \: = & \;
   c_{pd}\;  d \left[ \:\log( \theta_{s} ) \: \right]
  \: , \label{def_Betts_theta_s} \\
  q_d \:\: d\left( \frac{s}{q_d} \right)
 \: = \: 0  \: = & \;
  c_p \; d \left[ \:\log( \theta_{es} ) \: \right]
  \: , \label{def_Betts_theta_es} \\
  q_d \:\: d\left( \frac{s}{q_d} \right) 
  - \frac{L_{\mathrm{vap}}(T)}{T} \; dq_t
 \: = \: 0  \: = & \;
  c_p \; d \left[ \:\log( \theta_l ) \: \right]
  \: . \label{def_Betts_theta_l}
\end{eqnarray}
The specific heat $c_p = q_d\:c_{pd}+q_{sw}\:c_{pv}$ is a moist value in Eqs.(\ref{def_Betts_theta_es}) and  (\ref{def_Betts_theta_l}), with a corresponding moist value of the gas constant $R = q_d\:R_d+q_{sw}\:R_v$ involved in the terms $q_d\: d(s/q_d)$.
Equation~(\ref{def_Betts_theta_s}) can be solved exactly, since the result $\theta_s$ is given by Eq.(\ref{def_THs}).
The two other equations are solved with the approximations $c_p \approx c_{pd}$ and $R \approx R_d$ in \citet{Betts73} and \citet{BettsDugan73}.

The term $- \: (L_{\mathrm{vap}}/T) \: dq_t $ is equal to zero in Eq.(\ref{def_Betts_theta_l}), since $q_t$ is assumed to be a constant.
This term is further approximated by $- \: d\,[\, (L_{\mathrm{vap}}\: q_t) /T\,]$ and it corresponds to a differential of a certain constant term depending on $q_t$.
Therefore this constant term must  added to $c_p\:\ln(\theta_{es})$ in order to define $c_p\:\ln(\theta_l)$, and thus the symmetrical pair of potential temperature $\theta_{es}$ and $\theta_l$.
This precisely corresponds to the exponential term $\exp[\,-\, (L_{\mathrm{vap}}(T)\:q_t)/(c_{pd}\:T)\:]$ to be put in factor of $\theta_{es}$ to get $\theta_l$.
It is worthwhile noticing that the consequence of the approximations is that this exponential term is not a constant, since it depends on both $q_t$ and $T$, with the absolute temperature being a non-conservative variable.

Eqs~(\ref{def_Betts_theta_es}) and (\ref{def_Betts_theta_l}) do not correspond to changes in specific moist-air entropy $ds$, nor in the associated potential temperature ${\theta}_s$.
The equivalent and liquid-water potential temperature are defined by ${\theta}_{es} \approx \theta \: \exp[\:(L_{\mathrm{vap}} \: q_{sw})/(c_{pd}\:T)\:]$ and $\theta_l \approx \theta \: \exp[\:-\: (L_{\mathrm{vap}}\: q_l)/(c_{pd}\:T)\:]$, respectively.
They are approximately constant during pseudo-adiabatic processes undergone by a closed parcel.
The approximate features are due to $c_p \approx c_{pd}$, $R \approx R_d$ plus the hypotheses $(L_{\mathrm{vap}}/T) \:dq_{sw} \approx d(L_{\mathrm{vap}}\:q_{sw}/T)$ and $(L_{\mathrm{vap}}/T) \:dq_l \approx d(L_{\mathrm{vap}}\:q_l/T)$ recalled in \citet{Deardorff80}.

These approximations, together with the fact that $q_t$ is assumed to be a constant, explain the differences between $\theta_s$ and ${\theta}_{es}$ or $\theta_l$: only $\theta_s$ can be used as a conservative (isentropic) variable in case of open systems, because the additional terms depending on $q_t$ in $\theta_s$ given by Eq.(\ref{def_THs}) could not be derived from Eqs.(\ref{def_Betts_theta_es}) and (\ref{def_Betts_theta_l}), due to the arbitrary terms corresponding to the hypothesis $dq_t = 0$ and due to the aforementioned approximations needed to define ${\theta}_{es}$ or $\theta_l$.
There is no need to make these approximations for deriving the Third-Law-based moist-air entropy and $\theta_s$.

Similar ``conserved'' quantities depending on $q_t$ are included by hand in Eq.(4.5.15) of \citet{emanuel94}, to form another liquid-water potential temperature also denoted by $\theta_l$.
The ``conserved'' arbitrary quantity is equal to $(R_d+r_t\:R_v)\:\ln(p_0) - (R_d+r_t\:R_v)\:\ln(1+r_t/\epsilon) +r_t\:R_v\:\ln(r_r/\epsilon)$.
It is indeed a complex term that can only be justified by a wish to arrive at a result prescribed {\it a priori\/}.
Any other terms depending on $q_t$ and $r_t$ would lead to other possible conservative quantities different from $\theta_l$, making these manipulations unclear and questionable, in particular if $q_t$ is not a constant.

The important result obtained with the specific moist-air entropy expressed by (\ref{defS_THs}) is that $c_{pd} \: \ln( \theta_s )$ is an exact integral of (\ref{def_satur_adiabat}) expressed as $ds=0$ which corresponds to the Second Law property.
This result is obtained without manipulation of terms depending on $q_t$.
Therefore $\theta_s$ is a true measurement of the moist-air entropy and an exact generalization of $\theta_l$ or $\theta_e$.
New isentropic conservative properties can thus be observed for $\theta_s$ and for open systems, with changes in $q_t$ that must be balanced by changes in other variables, in order to keep $\theta_s$ unchanged.

    \subsection{The conservative properties of ${\theta}_{s}$: a synthesis}
    \index{sec25.7.3}\label{sec25.Thetas_Conservative}

\begin{table}
\caption{\em\small Analyses of several moist-air conservation properties for several moist-air potential temperatures.
\label{Table_conservative}}
\centering
\begin{tabular}{c|cccccc}
\hline
    List of moist-air properties  & $\theta_v$  &  $\theta_l$ &  $\theta_e$ &  $\theta'_e$  &  $\theta'_w$ &  $\theta_s$ \\ 
\hline
    unsat. adiab. / closed parcel / constant $q_v$  & (Y/N)$^{\text a}$ & A$^{\text b}$ & A & A  & A & Y$^{\text c}$ \\
\hline
 sat. adiabatic / closed parcel / constant $q_t$     &  (Y/N) & A & A & N$^{\text d}$ & N & Y \\
\hline 
 pseudo-adiab. / open parcel / $q_t=q_{sat}(T,p)$   & (Y/N) & I$^{\text e}$  & I & Y & Y & N \\
\hline
  isentropic / open parcel / varying $q_t$, $q_l$ or $q_i$  & (Y/N) & I & I & I & I & Y \\
\hline
\end{tabular}\\
$^{\text a}$Yes or No.
$^{\text b}$Approximate.
$^{\text c}$Yes.
$^{\text d}$No.
$^{\text e}$Irrelevant.
\end{table}

A synthetic view  is given in Table~\ref{Table_conservative}, where the moist-air conserved feature is tested for six potential temperatures (listed on the first line) and four moist-air conservative properties (described on the first column).
\vspace{-2.5mm}
\begin{itemize}
\item The virtual potential temperature $\theta_v = T_v \: (p/p_0)^\kappa$ is conserved for neutral buoyancy conditions.
Adiabatic or pseudo-adiabatic motions of closed or open parcels of moist-air do not automatically imply the conservation of $\theta_v$.
For these reasons, $\theta_v$ may be conserved, or not, independently of the adiabatic, pseudo-adiabatic or isentropic properties.
\item The Betts' liquid-water and equivalent potential temperature ($\theta_l$ and $\theta_e$) are both approximately conserved  for adiabatic motions of closed parcels of moist air, where $q_t$ is a constant.
Since it is an equivalent of the moist-air entropy, only $\theta_s$ is a true conserved adiabatic quantity and the approximate feature for both $\theta_l$ and $\theta_e$ is due to the two terms in Eq.(\ref{def_THs}) which depends on $T$ and $p$ and which are not included in $({\theta}_{s})_1$ given by Eq.(\ref{def_theta_s1_Betts}) (which behaves like $\theta_l$ if $q_t$ is a constant).
Both $\theta_l$ and $\theta_e$ are irrelevant for studies of open systems (like motions of real parcels of moist-air with varying $q_t$).
This is due to the method described in section~\ref{sec25.Thetas_adiab}, where $\theta_l$ and $\theta_e$ are derived from two (approximate) differential equations, with the assumption of constant $q_t$.
\item The two potential temperatures $\theta'_e$  and  $\theta'_w$ are defined in order to be conserved for pseudo-adiabatic processes only (section~\ref{sec25.Thetas_adiab}).
Since pseudo-adiabatic are incompatible with existing condensed water species, neither $\theta_l$ nor $\theta_e$ are conserved for adiabatic motions of closed parcels, and they are irrelevant for describing isentropic feature for open parcels (where contents in all water species may vary in time and in space).
Only approximate conservative feature are observed for $\theta'_e$  and  $\theta'_w$ for adiabatic motions of unsaturated parcels of moist-air, where it is of usual practice to define $\theta'_w$ by an adiabatic saturation process by imposing a constant value for the dry-air potential temperature $\theta$.
This can be explained because only $\theta_s$ is conserved during these moist-air adiabatic motions, and even if $q_t=q_v$ is a constant, the conservation of $\theta_s$ given by Eq.(\ref{def_THs}) does not mean that $\theta$ is conserved, due to the two terms in Eq.(\ref{def_THs}) which depends on $T$ and $p$ and which are not included in $({\theta}_{s})_1$ given by Eq.(\ref{def_theta_s1_Betts}) (which behaves like $\theta$ if $q_l=q_i=0$ and $q_t=q_v$ is a constant).
\item The moist-entropy potential temperature $\theta_s$ is not conserved by pseudo-adiabatic processes, since it is explained in section~\ref{sec25.Thetas_Pseudo_adiab} that the increase in $\theta_s$ is given by Eq.(\ref{def_s_pseud_adiab}).
$\theta_s$ is however interesting in that it is conserved for all other moist-air adiabatic processes (last column).
\end{itemize}
\vspace{-2mm}

The conclusion of this section is that pseudo-adiabatic processes must be studied by using $\theta'_e$ or $\theta'_w$, and that other adiabatic processes should be analyzed with the use of $\theta_s$, in order to avoid unnecessary approximations.
The well-known potential temperatures $\theta_l$ and $\theta_e$ are however relevant approximations of $\theta_s$ for studying conservative properties of closed parcels of moist-air.
Studies of isentropic feature for open systems (last line in Table~\ref{Table_conservative}) require the use of $\theta_s$ and they exclude the use of other moist-air potential temperatures.
As for the virtual potential temperature $\theta_v$, it is the specialized quantity suitable for studying impacts of buoyancy force and it cannot be used for analysing isentropic features.

 \section{\underline{Physical properties of the moist-air enthalpy and $T_h$}.} 
\index{sec25.8}\label{sec25.Physical_properties_H}

     \subsection{Applications for the moist-air enthalpy: turbulent surface fluxes}
     \index{sec25.8.1}\label{sec25.Businger_1982}

The problem of the need to manage relevant values for the reference enthalpies in atmospheric science  is explicitly addressed  in \citet{Businger82}.
The specific enthalpies are written as  $h_k - (h_k)_r = c_{pk}\: (T - T_r)$, where the index $k$ goes for dry air ($d$), water vapour ($v$), liquid water ($l$) and ice ($i$).
They can be written as $h_k = c_{pk}\:T + b_k$, provided that $b_k = (h_k)_r - c_{pk}\:T_r$.
The question asked in \citet{Businger82} is to know whether or not it is important to determine the numerical values of $b_k$ -- and thus the reference values $(h_k)_r $ --  in order to compute the vertical turbulent fluxes of the moist-air enthalpy $h$.

It is a usual practice in atmospheric science to set reference enthalpies of dry air equal to zero for a given reference temperature, typically at the high temperature $T_r = 273.15$~K~$=0^{\circ}$C.
The choice of zero-enthalpies for water species are more variable.
They are set to zero either for the water vapour or for the liquid water enthalpies, for the same reference temperature $0^{\circ}$C and with the latent heats obviously connecting the other water-enthalpies by $L_{vap} (T_r)= (h_v)_r - (h_l)_r$ or $L_{fus} (T_r)= (h_l)_r - (h_i)_r$.

It is shown in \citet{Businger82} that it is the choice of zero-enthalpies for both dry air and liquid water (at $T_r = 0^{\circ}$C) that is in agreement with well-established procedures for computing surface turbulent fluxes (the ones still retained nowadays).
Otherwise, additional fluxes of $q_t$ should be added to the moist-air enthalpy flux, leading to other definitions of this turbulent flux.
The same hypothesis $(h_d)_r = (h_l)_r = 0$ is retained in the review of Fuehrer and Friehe (2002) and for $T_r = 0^{\circ}$C.

It is however unlikely that such arbitrary fluxes of $q_t$ may be added or cancelled to the enthalpy flux, leading to arbitrary closure for the computation of turbulent fluxes of moist air.
The same is true for the vertical integral and the horizontal or vertical gradients of $h$, for which terms depending on $q_t$ could be of a real importance if $q_t$ is not a constant.

The accuracy and the relevancy of the hypotheses $(h_d)_r = (h_l)_r$ ($=0$?),  $(h_d)_r = (h_v)_r$ ($=0$?), $b_v = (h_v)_r - c_{pv}\:T_r = 0$ or $b_l = (h_l)_r - c_l\:T_r = 0$ are analyzed in the next section.

     \subsection{Impacts of Trouton's rule and of the coincidence $(h_l)_r \approx (h_d)_r$}
     \index{sec25.8.2}\label{sec25.Trouton}

The analyses of the ``constant heat capacity'' curves ($cst$ straight lines) of enthalpies depicted in Fig.~\ref{fig25.4}(b) can be used to test 
the three kinds of assumptions listed in \citet{Businger82}:
\vspace{-2.5mm}
\begin{itemize}
\item $h_l(T_r) = h_d(T_r)$ for some reference temperature $T_r>150$~K; 
\item $h_l(T_r) = 0$ and $ h_d(T_r) = 0$ for some reference temperature $T_r>150$~K; 
\item $h_l(T_r) = 0$ and $ h_d(T_r) = 0$ for the extrapolated limit $T_r = 0$~K, with the consequence of  linear laws $h_v(T) =  c_{pv}  \: T$ and $h_l(T) =  c_l \: T$ valid for temperature $T>150$~K.
\end{itemize}
\vspace{-2mm}

The reference temperature for which $h_l(T_r) = h_d(T_r)$ can be computed with $h^0_l$ and $ h^0_d$ given by  (\ref{def_h0l}) and (\ref{def_h0d}) and with the properties $h_l(T_r)=h_l(T_0)+c_l\:(T_r-T_0)$ and $h_d(T_r)=h_d(T_0)+c_{pd}\:(T_r-T_0)$. 
It is equal to $T \approx 241.4 $~K$ \; = -31.75 \:^{\circ}$C (the grey spot in Fig.~\ref{fig25.4}(b)).
It is clear from this Fig.~\ref{fig25.4}(b) that the differences between $h_l(T)$ and $h_d(T)$ become larger and larger for increasing temperatures.
In particular, for $30 \:^{\circ}$C $h_l \approx 760$~kJ~kg${}^{-1}$ is $36$~\% larger than $h_d \approx 560 $~kJ~kg${}^{-1}$.

It is therefore only a coincidence and a crude approximate property that the dry-air and liquid-water thermal enthalpies are close to each other for the atmospheric range of temperature.
Except for the too low reference temperature $T_r = -31.75 \:^{\circ}$C, this results  invalidates the hypothesis recalled in section~\ref{sec25.Reference_values}, where the property $(h_l)_r = (h_d)_r$ is often assumed to cancel the second term in the second line of (\ref{def_h7}).
It also demonstrates that additional terms should be added to the usual turbulent fluxes of enthalpy, according to the conclusions of \citet{Businger82}.

Values of $h^0_i$ are about half those of $h^0_l$, and due to the Trouton's rule, values of $h^0_v$ are much higher than those of $h^0_i$ and $h^0_l$.
Trouton's rule - see the review of \citet{Wisniak01} -  states that the general property $L_{vap}/\:T \approx 88$~J~mol${}^{-1}$~K${}^{-1}$ holds true for almost all  substances at their boiling temperature.
Since boiling temperature for H${}_2$O (373.15~K) is about $4.4$ times greater than the ones for N${}_2$ (77.4~K) and  O${}_2$ (90~K), the latent heats of sublimation and vaporisation of water are thus logically the dominant features in Fig.~\ref{fig25.4}.
This invalidates the possibility $(h_v)_r = (h_d)_r$ for which the second term in the  second line of (\ref{def_h6}) may be discarded.
Even if $(h_v)_r - (h_d)_r$ can be written as $(h_l)_r - (h_d)_r + L_{\mathrm{vap}}(T_r)$, with the dominant term $L_{\mathrm{vap}}(T_r)$ leading to possible correction values in (\ref{def_h6}), the remaining term $(h_l)_r - (h_d)_r$ is different from zero in (\ref{def_h7}) and must be taken into account.

Moreover, since the liquid water and water vapour straight lines depicted on Fig.~\ref{fig25.4}(b) above $150$~K do not intersect the origin (if these straight lines were continued toward $T=0$~K), this does not confirm the properties $b_v=0$ or $b_l=0$, namely the linear laws $(h_v)_r =  c_{pv}  \: T_r$ or $(h_l)_r =  c_l \: T_r$, which could allow the cancellation of the second lines of  (\ref{def_h8}) and (\ref{def_h9}) whatever values of $q_t$ may be.

     \subsection{Comparisons of moist-air enthalpy with various moist-static energy formulae}
     \index{sec25.8.3}\label{sec25.comparisons_h_Th}

It is recalled in this section that the concept of Moist Static Energy (MSE) is intimately linked with the conservation of the moist-air entropy ($s$), and with the computation of the moist-air enthalpy ($h$).

The link between the moist-air enthalpy and entropy is due to the Gibbs equation 
\begin{align}
  T \: \frac{ds}{dt}  & = \:
   \frac{dh}{dt} 
   \: - \: \frac{1}{\rho} \frac{dp}{dt} 
   \: - \: \sum_k {\mu}_k \: \frac{d q_k}{dt} 
  \:  , \label{def_Gibbs_1} 
 \end{align}
where the local Gibbs function is defined by ${\mu}_k = h_k - T \: s_k$ and where the sum over the index $k$ goes for dry air ($d$), water vapour ($v$), liquid water ($l$) and ice ($i$).

The reason why it is possible to associate the moist-air enthalpy $h$ with $s$ is that, for stationary pure vertical motions, the material derivative are equal to $d/dt = w \: {\partial}/{\partial z}$, with the vertical velocity $w$ factorizing all the terms in (\ref{def_Gibbs_1}) and which can be omitted.
For vertical hydrostatic motions $- \: {\rho}^{-1}\:{\partial p}/{\partial z} = {\partial \phi}/{\partial z}$.
If the parcel is closed (constant $q_t$) and undergoes reversible adiabatic processes (no super-cooled water, but with possible reversible  condensation,  evaporation, fusion, ... processes), then ${\partial s}/{\partial z} = 0$ and $\sum_k \: {\mu}_k \:{\partial q_k}/{\partial z} = 0$.
The Gibbs equation can then be written as
\begin{align}
  T \: \frac{\partial s}{\partial z}  
  & =  \: 0  \: =  \: 
   \frac{\partial ( h + \phi)}{\partial z} 
  \:  . \label{def_Gibbs_MSE} 
 \end{align}

The quantity $h + \phi$  is thus a conserved quantity for vertical motions and provided that all the previous assumptions are valid (adiabatic and hydrostatic vertical motion, closed parcel).
This sum $h + \phi$  is called the generalized enthalpy in \citet{ambaum10}.

The consequence of the adiabatic conservative property (\ref{def_Gibbs_MSE}) is that the generalized enthalpy can be used to represent the convective saturated updrafts.
There is thus a need to compute $h+\phi$ at the bottom of an updraft, and therefore to compute the moist-air enthalpy $h$ itself, before using the property $h+\phi = Cste$ to determine the local properties of this updraft ($T$, $q_v$, $q_l$, $q_i$) at all levels above the cloud base.

It is expected that the moist-air specific thermal enthalpy $h$ given by Eq.(\ref{def_h_b}) is the quantity that enters the generalized enthalpy $h+\phi$ in Eq.(\ref{def_Gibbs_MSE}).
It is however a common practice in studies of convection to replace $h+\phi$ by one of the MSE quantities  defined by
\begin{align}
\mbox{MSE}_d \: & = \: 
{c}_{pd} \: T \: + \:  L_{\mathrm{vap}}\:q_v \: + \:  \phi 
  \: , \label{def_MSEd} \\
\mbox{MSE}_l \: & = \: 
{c}_{pd} \: T \: - \:  L_{\mathrm{vap}}\:q_l \: + \:  \phi 
  \: , \label{def_MSEl} \\
\mbox{LIMSE} \: & = \: 
{c}_{pd} \: T \: - \:  L_{\mathrm{vap}}\:q_l \: -  \:  L_{\mathrm{sub}}\:q_i \: + \:  \phi 
  \: . \label{def_LIMSE}
\end{align}
The formulation MSE$_d$ is the more popular one.
It is considered in \citet{AS74} as ``approximately conserved by individual air parcels during moist adiabatic processes''.
It is mentioned in \citet{Betts75} that it is approximately an analogue of the equivalent potential temperature $\theta_e$.
It is used as a conserved variable for defining saturated updraughts in some deep-convection schemes \citep{Bougeault85}.

A liquid-water version MSE$_l$ is defined in \citet{Betts75} by removing the quantity $L_{\mathrm{vap}}\:q_t$ (assumed to be a constant) from (\ref{def_MSEd}), leading to (\ref{def_MSEl}).
It is considered in \citet{Betts75} that MSE$_l$ is an analogue of the liquid-water potential temperature $\theta_l$.
The formulation MSE$_l$ is generalized in \citet{KR03} and in \citet{Breth_al_05} by removing a term $L_{\mathrm{sub}} \:q_i$ to (\ref{def_MSEl}), for sake of symmetry, leading to the liquid-ice static energy LIMSE given by (\ref{def_LIMSE}), which is used as a conserved variable for defining saturated updraughts in some shallow-convection schemes \citep{Bechtold05}.

It is shown in \citet{marquet15} that the last terms in factor of $q_t$ in Eq.(\ref{def_h_b}) can be roughly approximated by the dominant term $L_{\mathrm{vap}}(T)$.
This corresponds to the coincidence $h^0_d \approx h^0_l$  and to the Trouton's rule described in Section~\ref{sec25.Trouton}.
Accordingly, the generalized moist-air thermal enthalpy can be approximated by $h   + \phi \: \approx \: h_{ref} \: + \:$FMSE, where
\begin{align}
\mbox{FMSE} & = \: 
{c}_{pd} \: T
 \: + \:
   L_{\mathrm{vap}}\:q_v  
\: - \:  
   L_{\mathrm{fus}}\:q_i
\: + \: \phi
  \: . \label{def_h_approx_a}
\end{align}
This formulation correspond to the frozen moist static energy denoted by $s$ in Eq.(1.25: to be check!) of Section 1.4.3 of this Convection Book.
This quantity FMSE is defined by and studied in \citep{Breth_al_05,Rooy_al_13}.
It is equal to the sum of LIMSE plus $L_{\mathrm{sub}} \:q_t$ and it is equal to MSE$_d$ for positive Celsius temperatures ($q_i=0$).
The drawback of this approximate formula is that it does not possess the same symmetry as in Eq.(\ref{def_h_b}), where $L_{\mathrm{vap}}$ is logically replaced by $L_{\mathrm{sub}}$ if $q_l$ is replaced by $q_i$.
This means that the approximation used to establish Eq.(\ref{def_h_approx_a}) is probably not accurate enough.
 
MSE is sometimes defined with  $c_{pd}$ replaced in (\ref{def_MSEd}) by the moist value $c_p$, leading to
\begin{align}
\mbox{MSE}_m \: & = \: 
{c}_p \: T \: + \: L_{\mathrm{vap}}\:q_v \: + \:  \phi 
  \: , \label{def_MSE} 
\end{align}
This version is used in some convective schemes \citep{Gerard09}.

Differently, MSE quantities are often expressed per unit mass of dry air and with the assumptions $h^0_d=h^0_l=0$, like the moist enthalpy $k$ in \citet{emanuel94} which is equivalent to the MSE defined by $h^{\star}_v$ in \citet{ambaum10}.
\begin{align}
h^{\star}_v  \:  & = \: 
[ \: {c}_{pd} \: + q_t \: (c_l-{c}_{pd}) \: ] \: T \: + \:  L_{\mathrm{vap}}\:q_v \: + \: \phi 
  \: . \label{def_hv_A10} \\
k + \phi / q_d \: = \: h^{\star}_v / q_d \: & = \: 
({c}_{pd} \: + r_t \: c_l) \: T \: + \:  L_{\mathrm{vap}}\:r_v \: + \: (1+r_t) \: \phi 
  \: . \label{def_kv_E94}
\end{align}

\begin{figure}[ht]
\centering
\includegraphics[width=0.65\linewidth,angle=0,clip=true]{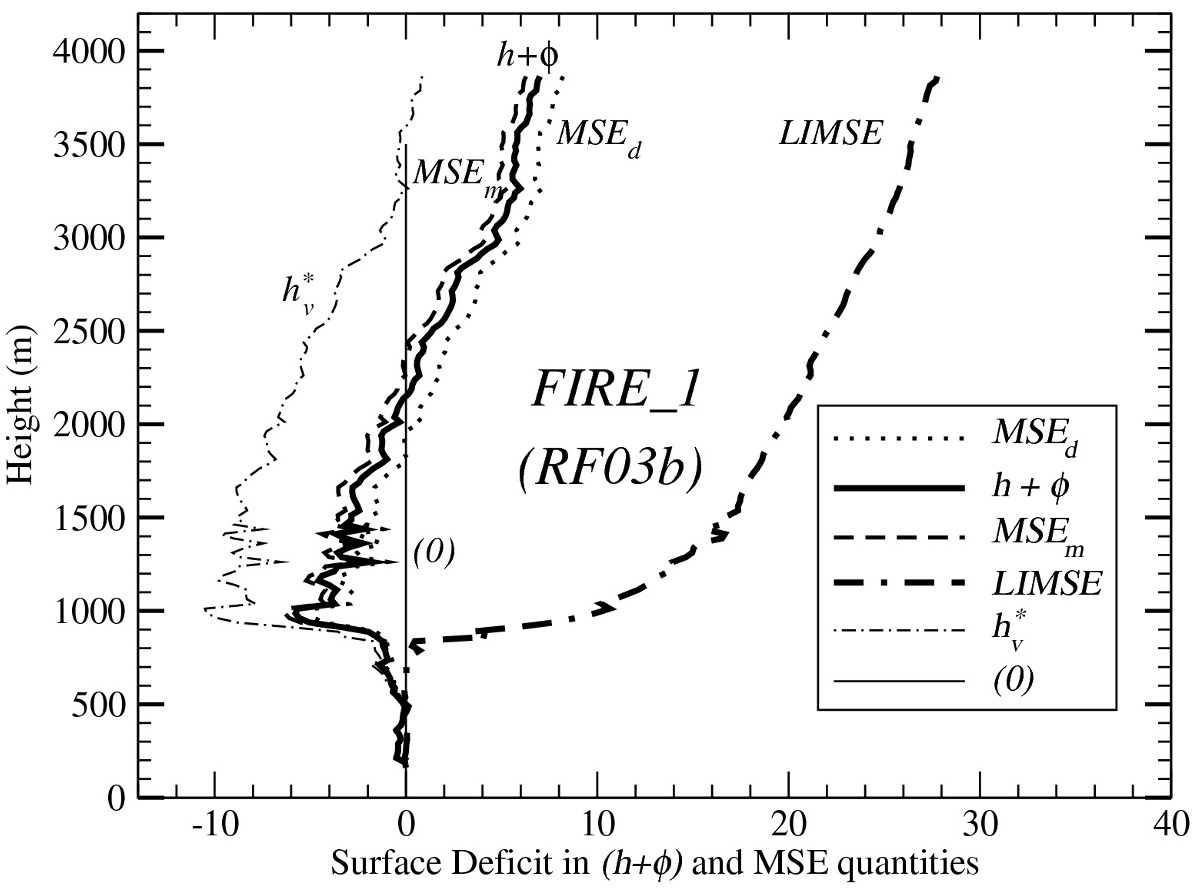}
\includegraphics[width=0.65\linewidth,angle=0,clip=true]{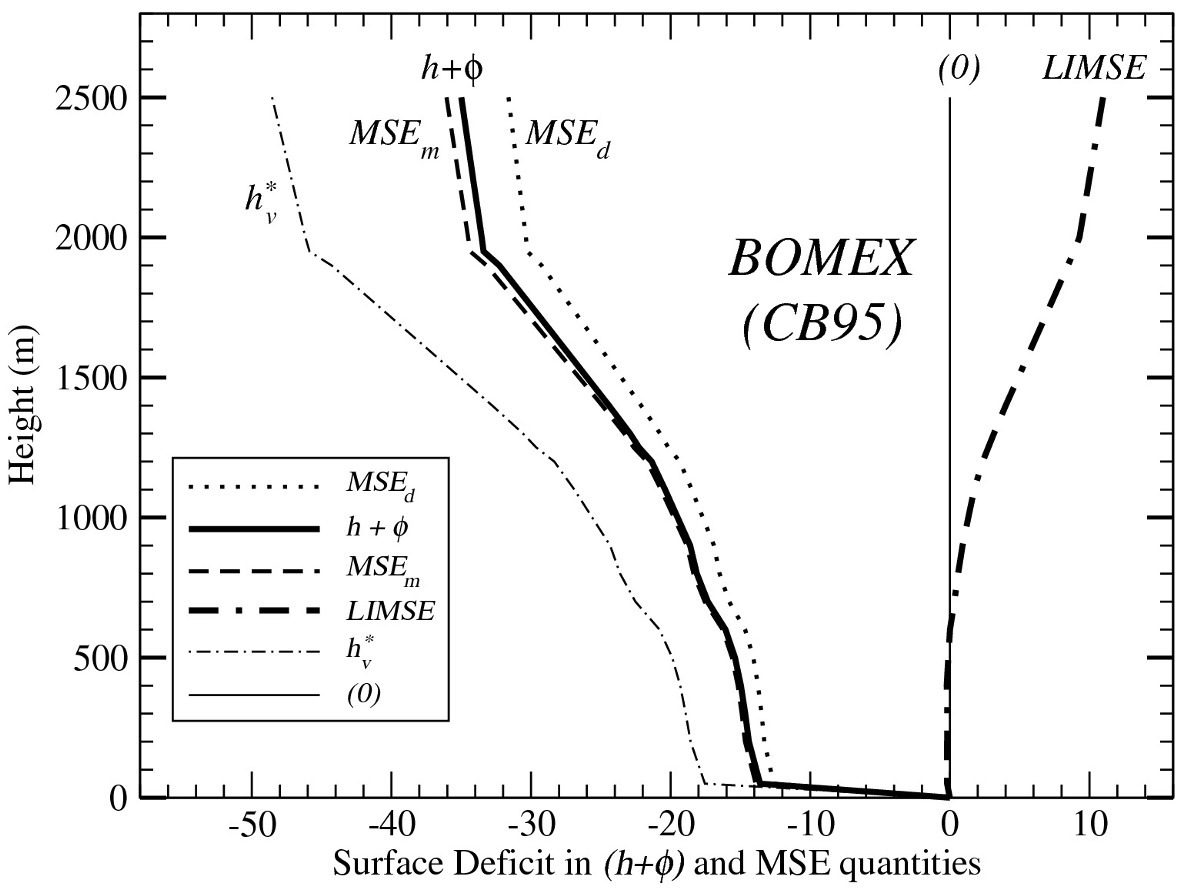}
\caption{\it \small Surface deficit in generalized enthalpy $h+\phi$ and in MSE quantities for the FIRE-I (RF03b) stratocumulus and the BOMEX shallow cumulus datasets. 
Units are in kJ~kg${}^{-1}$.}
\label{fig25.7}
\end{figure}

The aim of this section is thus to compare the generalized enthalpy $h+\phi$, with $h$ given by Eq.(\ref{def_h_b}), with some of the MSE quantities given by Eqs.(\ref{def_MSEd})-(\ref{def_kv_E94}).
The properties of surface-deficit charts are of common use in studies of convective processes and
two of them are depicted on Fig.~\ref{fig25.7}, for the FIRE-I (RF03B) and BOMEX datasets.

It is shown that the two formulations MSE$_m$ and MSE$_d$ ($=$FMSE) are close to $h+\phi$ at all levels for both cases, with the generalized enthalpy located in between the others and with MSE$_m$ being a better approximation for $h+\phi$.
The generalized enthalpy is more different from the other quantities LIMSE and $h^{\star}_v$ (with $k + \phi/q_d$ very close to $h^{\star}_v$, not shown).
This means that LIMSE and $h^{\star}_v$ cannot represent accurately the generalized enthalpy $h+\phi$.

Large jumps in all variables are observed close to the surface for BOMEX, except for LIMSE, due to the impact of large values of surface  specific humidity that are not taken into account in LIMSE.
The impact of surface values are not observed for FIRE-I, since the air-flight measurements were realized above the sea level.

The conclusion of this section is that MSE$_m$ is probably the best candidate for approximating $h+\phi$.
However, the fact that systematic differences between $h+\phi$ and MSE$_m$ exist in the moist lower PBL only, and not in the dry-air above, may have significant physical implications if the purpose is to analyze moist enthalpy budgets accurately, or differential budgets of it, or to understand the convective processes (entrainment and detrainment), or to validate the long term budgets for NWP models and GCMs by comparing them with climatology or reanalyses.
For these reasons, it is likely important to define the moist-air enthalpy $h$ by Eq.(\ref{def_h_b}) if the aim is to  use conservative MSE quantities based on the generalized enthalpy.

 \section{\underline{Summary and discussion of possible new applications}.}
\index{sec25.9}\label{sec25.Conclusion}

It is shown in this Chapter~25 that it is possible to compute directly the moist-air entropy expressed by $s =  s_{ref} + c_{pd} \: \ln ( \theta_s)$, in terms of the potential temperature $\theta_s$ given by Eq.(\ref{def_THs}).
The moist-air thermal enthalpy can be similarly expressed by $h =  h_{ref} + c_{pd} \: T_h$, in terms of the enthalpy temperature $T_h$ given by Eq.(\ref{def_Th}).

It is shown in previous sections that these new formulas are associated with the First, Second and Third Laws of thermodynamics and that they are derived with a minimum of hypotheses.
The important result is that these formulas offer the opportunity to analyze new atmospheric processes, since it is possible to evaluate if isentropic conditions prevail within those regions where moist-air must be considered as an open system, namely where $q_t$ is varying in time and in space.
In particular, computations of change in moist-air entropy are suitable and relevant for convective processes, simply by examining the distribution of $\theta_s$ in those regions where entrainment, detrainment, precipitations or evaporation take place.
These studies cannot be realized by using other meteorological potential temperatures.

A first example of possible application of moist-air entropy to convective processes concerns the top-PBL entrainment of marine stratocumulus, where the top-PBL jump in moist-air entropy seems to control the turbulent and convective processes, by maintaining or evaporating the cloud depending on the sign of the jumps in $\theta_s$.
This result is not observed with the sign of top-PBL jumps in $\theta_l$ or $\theta_e$ that cannot directly explain the instability of marine stratocumulus, simply because $\theta_l$ and $\theta_e$ cannot represent the moist-air entropy if $q_t$ is not a constant.

A second example is given by theories of conditional symmetric instability and slantwise convection.
These theories are expressed in \citet{BH79} in terms of the moist-air Brunt-V\"ais\"al\"a frequency and the vertical gradient of $\theta'_w$.
The wet bulb potential temperature $\theta'_w$ is replaced by $\theta_e$ in subsequent studies  \citep{Emanuel83a, Emanuel83b, Emanuel85, THEM85,EFT87, emanuel94}.
Both $\theta'_w$ and $\theta_e$ are used as equivalent of the moist-air entropy, with $d\theta_e/dt \neq 0$ only for diabatic processes other than latent heat release.
It is however irrelevant to use $\theta_e$ in those regions where $q_t$ is not a constant.
This is a clear motivation for replacing the use of $\theta_e$ by the use of $\theta_s$ for studying slantwise convection.

A third example concerns the change in moist-air entropy which is is a key quantity entering Emanuel's vision of a tropical cyclone as a Carnot heat engine   \citep{Emanuel86, Emanuel91, Holand97, Camp_Mont_2001, Bryan_Rotunno_09, Tang_Emanuel_2012}.
It is mostly the quantity $\theta_e$ which is used to compute the difference between the entropies $\Delta s$ of moist air in the ambient environment and near the storm centre.
However, $q_t$ is rapidly varying between these regions.
This is a motivation for replacing the use of $\theta_e$ by the use of $\theta_s$ for studying hurricanes, typhoons or cyclones.

Even if it is valuable to continue to analyze pseudo-adiabatic processes, if needed, with the help of constant values of $\theta'_e \approx \theta_e$ or $\theta'_w$ (see the Table~\ref{Table_conservative}), the three aforementioned examples are typical applications where $\theta_l$ or $\theta_e$ must be replaced by $\theta_s$ in order to compute and compare values of moist-air entropy in regions with varying $q_t$.

The first example demonstrated in \citet{marquet11} and recalled in section~\ref{sec25.Sc_to_Cu} concerns the constant feature of $\theta_s$ in the moist-air PBL of marine stratocumulus.
The second example dealing with the moist-air potential vorticity and the slantwise-convection criteria is documented in \citet{marquet14} and in section~\ref{sec25.moist_PV}.
The third example has been documented by analyzing drop-sounding and LAM outputs for several cyclones (unpublished results).

\begin{figure}[ht]
\centering
\includegraphics[width=0.95\linewidth,angle=0,clip=true]{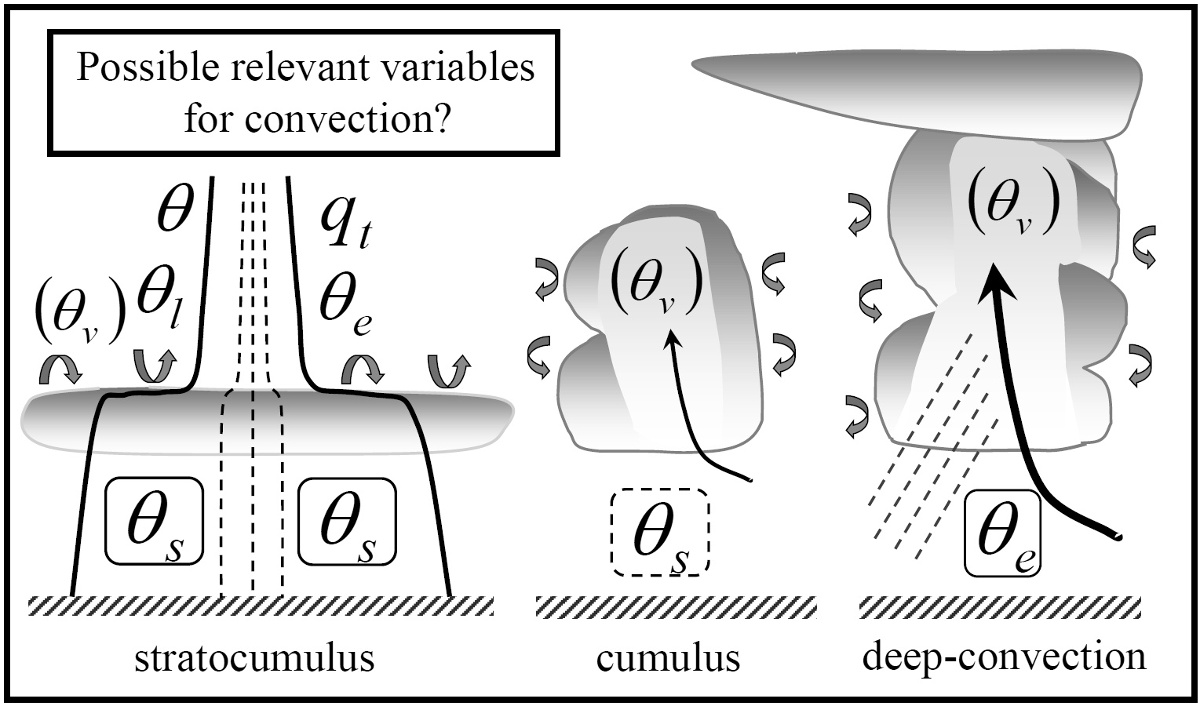}
\caption{\it \small A list of relevant variables for different regimes of convection.}
\label{fig25.9}
\end{figure}

Three classes of convective phenomena are plotted in Fig.~\ref{fig25.9}.
From the left to the right are depicted: stratiform clouds (stratocumulus), shallow convection (non-precipitating cumulus) and deep convection (precipitating cumulus or cumulonimbus).

One of the goal of this figure is to recall that the study of $\theta_v$ is relevant for all clouds, since it is the relevant parameter for the buoyancy force.
Therefore, the conservative properties observed for $\theta_v$ are not related to the Second Law and they do not correspond to the moist-air entropy.

It is recalled that $\theta_e$ (in fact $\theta'_w$ or $\theta'_e$) is conserved during pseudo-adiabatic processes and is thus relevant for describing the precipitating saturated updrafts in deep-convection clouds (provided that the existing non-precipitating cloud condensed water species are discarded).

However, $\theta_e$ is not suitable for describing the non-precipitating parts of the deep-convection clouds (since closed parcels implies constant $q_t$ and are not compatible with pseudo-adiabatic processes), nor the impact on moist-air entropy of the flux of matter occurring at the edge of clouds (since entrainment and detrainment are not pseudo-adiabatic processes and since they act on open systems, where changes in $q_t$ are large and rapid).

It is recalled that a stratocumulus exhibits large top-PBL jumps in $\theta$, $\theta_l$, $\theta_e$ or $q_t$.
Only $\theta_s$ remains close to neutrality (both within the PBL and the entrainment region, with a continuous variation with the free-air located above).
For these clouds, the entropy potential temperature $\theta_s$ should replace $\theta_l$ and $\theta_e$, since $q_t$ is not a constant and since that $\theta_l$ and $\theta_e$ cannot represent the moist-air entropy of such open systems.

An important problem concerns the shallow convection, where no conserved variable really exists.
Since the condensed water are not withdrawn by the precipitations, pseudo-adiabatic variables cannot represent the updrafts of non-precipitating cumulus.
If adiabatic motion of closed parcels are considered, namely if isentropic processes are imagined, then the use of $\theta_s$ should replace those of $\theta_l$ and $\theta_e$, according to the conclusion of the  Table~\ref{Table_conservative}.
Moreover, only $\theta_s$ remain relevant for studying the impact of lateral entrainment and detrainment processes, due to the edges of clouds which can be considered as open systems.

However, the important limitation for the possibility that $\theta_s$ may be a conserved quantity for shallow convection is that the diabatic impact of radiation is important for these clouds.
It could be useful to interpret the problem differently: namely to analyze the non-equilibrium state of the cloud in terms of the diabatic impacts on moist-entropy and $\theta_s$, rather than to search for a adiabatic magic quantity which may not exist.

The common feature for the three kind of clouds is that it is always possible to compute the moist-air entropy in terms of $\theta_s$, whatever diabatic or adiabatic processes, and closed or open parcels, are considered.
Except for very specialized cases (pseudo-adiabatic or closed adiabatic parcels), it should be worthwhile to analyze the conservative feature of $s = s_{ref} + c_{pd} \: \ln(\theta_s)$ within clouds, at the edges of clouds and outside the clouds, automatically leading to an isentropic states for constant values of $s$ (whatever changes in $q_t$ may be), or to diabatic heating production of moist-air entropy given by Eq.(\ref{def_s_pseud_adiab}) for varying values of $s$.

It is explained in section~\ref{sec25.comparisons_h_Th} that there is an alternative description of conservative properties of clouds in terms of the moist-air generalized enthalpy $h+\phi$ defined by Eq.(\ref{def_Gibbs_MSE}), with the thermal enthalpy $h$ given by Eq.(\ref{def_h_b}).
The validity of the equation $\partial s /  \partial z = \partial (h + \phi) /  \partial z = 0$ relies on several assumptions: vertical, adiabatic and stationary motion of a closed parcel.
Since the Second law is invoked via the use of $\partial s /  \partial z = (c_{pd}/\theta_s) \: \partial \theta_s /  \partial z =0$, the conservation of $h + \phi$ is equivalent to the conservation of $\theta_s$, and approximately of  $\theta_l$ and $\theta_e$ since $q_t$ is assumed to be a constant.
This {\it a priori\/} excludes the use of $h + \phi$ in pseudo-adiabatic regions and suggests the use of $\theta_s$ instead  of $\theta_e$ or $\theta'_e$.

It is explained in section~\ref{sec25.comparisons_h_Th} that $h+\phi$ can be approximated with a good accuracy by MSE$_m$, MSE$_d$ or FMSE, but not by any of MSE$_l$, LIMSE, $k+\phi/q_d$ or $h^{\star}_v$.
Moreover, it is likely that systematic differences between the generalized enthalpy and MSE quantities are important enough to justify the use of the more general formula $h+\phi$  valid for the moist-air thermal enthalpy given by Eq.(\ref{def_h_b}).

The fact that it is possible to compute the entropy of matter in the atmosphere by the Third-Law based formula may be used for addressing in a new way both computations of budget of moist-air entropy and the assessment of  ``maximum entropy production''  (or MEP\footnote{
It is quoted in \citet{Ozawa_al_2003} that ``it was \citet{Paltridge75,Paltridge78} who first suggested that the global state of the present climate is reproducible, as a long-term mean, by a state of maximum Maximum Entropy Production'' ($\: = \:$ MEP).
The use of variational methods associated with the MEP principle can be solved by using principle of least action.
The relevance of such MEP principles for the atmosphere is described, analyzed and discussed by many authors
 \citep{NicolisNicolis80,Jaynes_80,Paltridge81,Grassl_81,Stephens_OBrien_1993,
   Nicolis99,Paltridge2001,Dewar_2003,Nicolis2003,Kleidon_al_2006,Kleidon_09,
   Goody07,Paltridge_al_2007,
   NicolisNicolis2010, Seleznev_Martyushev_2011,
   Pascale_al_2011bis,
   Ebeling-Feistel_2011,Martyushev_Seleznev_2013,
   Martyushev_Entropy_2013}.
}) principle, for either the global atmosphere or for local turbulent or convective processes.

Indeed, the change in moist-air entropy $ds/dt = - \overrightarrow{\nabla} . \, \vec{J}_s \: + \: \sigma_s$ is usually computed via the right-hand side flux ($\vec{J}_s$) and positive production rate ($\sigma_s$) of moist-air entropy \citep{Peixoto_al_1991,Stephens_OBrien_1993,Pauluis_Held_02,Pascale_al_2011}.
The aim of these studies is to analyze if the production of entropy by the moist-air atmosphere, lands and oceans can balance the average sink of about $1$~W~K${}^{-1}$~m${}^{-2}$ imposed by the imbalanced of incoming low-value solar and outgoing high-value infrared radiation entropies, including impact studies of local turbulent or convective processes.
The novelty offered by the Third-Law based entropy is that it is possible to compute directly the left-hand side of the moist-air entropy equation $ds/dt$ by using $s = s_{ref} + c_{pd} \: \ln(\theta_s)$ and a finite difference formulation $[\:s(t+dt)-s(t)\:]/dt$, leading to new comparisons  of NWP models or GCMs with analyzed or observed values.

The Third-Law based entropy can also be used to study the MEP principle in a new way.
It is explained in sections~\ref{sec25.comparisons_theta_sle} and \ref{sec25.Sc_to_Cu} that $\theta_s$ is about in a $2/3$ position between $\theta \approx \theta_l$ and $\theta_e \approx \theta'_e$.
This means that the moist-air entropies computed in \citet{emanuel94,Pauluis_al_10,Pauluis_11} are underestimated or overestimated when they are computed with $\theta_l$ or $\theta_e$, respectively.
Therefore, the moist-air entropy productions based on $\theta_l$ or $\theta_e$ must be different from those based on $\theta_s$, especially in regions with high values of $q_t$ like convective regions, storms or cyclones.
For these reasons, a MEP principle based on the moist-air entropy expressed in terms of $\theta_s$ could be more relevant than those based on the dry-air entropy (namely $\theta$ or $c_{pd}\:T + \phi$).
From $\Lambda_r = [ \: (s_v)_r - (s_d)_r \: ]/c_{pd}$, the moist-air entropy $s(\theta_s)$ depends on reference entropies determined from the Third Law, and thus on experimental values.
Therefore $s(\theta_s)$ can be obtained neither from the Shannon information entropy $-\sum_i p_i \log(p_i)$ nor from the relative (Kullback and Leibler) information entropy $\sum_i p_i \log(p_i/q_i)$, two theoretical quantities which can only deal with one-component systems like dry air, but not with the moist air.


 \section{\underline{List of symbols and acronyms}.}
\index{sec25.6}\label{sec25.Symbols}

\begin{tabbing}
 -------------\=  -------------------------------------- --\= \kill
 $a$ \>  a weighting factor in \citet{Pauluis_al_10} \\
 $b_k$ \>  $(h_k)_r - c_{pk}\:T_r$  for species $k$ in \citet{Businger82} \\
 $C$ \>  a transition parameter  \citep{marquet_geleyn13} \\
 $c_{pd}$ \> specific heat for dry air   \>($1004.7$~J~K${}^{-1}$~kg${}^{-1}$) \\
 $c_{pv}$ \> specific heat for water vapour \>($1846.1$~J~K${}^{-1}$~kg${}^{-1}$) \\
 $c_{l}$  \> specific heat for liquid water \>($4218$~J~K${}^{-1}$~kg${}^{-1}$) \\
 $c_{i}$  \> specific heat for ice          \>($2106$~J~K${}^{-1}$~kg${}^{-1}$) \\
 $c_p$ \> specific heat at constant pressure for moist air, \\
       \> $ = \: q_d \: c_{pd} + q_v \: c_{pv} + q_l \: c_l + q_i \: c_i  $ \\
 $c_p^{\star}$ \> specific heat at constant pressure depending on $r_t$ \\
 $c_v$ \> specific heat at constant volume for moist air \\
 $(d,v,l,i)$  \> lower-scripts for dry-air, water vapour, liquid water and ice  \\
 $D_C$ \>  a moist-air parameter in \citet{marquet_geleyn13} \\
 $\Delta H^0_f$  \> the standard enthalpies of formation \\
 $\Delta H^0_r$  \> the standard enthalpies of reaction \\
 $\Delta_2$  \> the CTEI line in the ($q_t$, $\theta_l$)  diagram \\
 $\Delta(q_t)$  \> top-PBL jump in $q_t$ for stratocumulus (and $\Delta(\theta_l)$, $\Delta(\theta_s)$, ...)\\
 $\eta$   \> $=1+\delta =R_v/R_d \approx 1.608$ \\
 $\varepsilon$ \> $=1/\eta=R_d/R_v \approx 0.622$ \\
 $\kappa$ \> $=R_d/c_{pd}\approx 0.2857$ \\
 $\gamma$ \> $= \eta \: \kappa \ = R_v/c_{pd} \approx 0.46$ \\
 $\lambda$ \> $= c_{pv}/c_{pd}-1 \approx 0.8375$ \\
 $e$       \> the water-vapour partial pressure \\
 $e_{sw}(T)$ \> partial saturating pressure over liquid water \\
 $e_{si}(T)$ \> partial saturating pressure over ice \\
 $e_r$      \> the water-vapour reference partial pressure,\\
            \> with $\: e_r = e_{ws}(T_r=T_0) \approx 6.11$~hPa \\
 $e_0$     \> the water-vapour standard partial pressure ($1000$~hPa)\\
 $e_i$      \> the internal energy ($=h-p/\rho=h-R\:T$)\\
 $\phi$ \>  the gravitational potential energy ($=g\: z + \phi_0$ )    \\
 $\Gamma$              \> the moist-air adiabatic lapse-rate \\
 $\Gamma_{sw}$    \> the liquid-water saturating moist-air adiabatic lapse-rate \\
 $F(C)$ \>  a parameter in \citet{marquet_geleyn13} \\
 $g$        \> $9.8065$~m~s${}^{-2}$ magnitude of gravity \\
 $h$        \> the moist-air specific enthalpy \\
 $h_{d}$  \>  enthalpy of dry air\\
 $h_{v}$  \>  enthalpy of water vapour\\
 $h_{l}$  \>  enthalpy of liquid water\\
 $h_{i}$  \>  enthalpy of ice\\
 $h_{ref}$  \>  a reference value for enthalpy \\
 $(h_{d})_r$  \> reference enthalpy of dry air at $T_r$\\
 $(h_{v})_r$  \> reference enthalpy of water vapour  at $T_r$\\
 $(h_{l})_r$  \> reference enthalpy of liquid water  at $T_r$\\
 $(h_{i})_r$  \> reference enthalpy of ice  at $T_r$\\
 $h^0_d$   \> standard specific enthalpy of the dry air ($530$~kJ~kg${}^{-1}$ )\\
 $h^0_v$   \> standard specific enthalpy of the water vapour ($3133$~kJ~kg${}^{-1}$) \\
 $h^0_l$   \> standard specific enthalpy of the liquid water ($632$~kJ~kg${}^{-1}$) \\
 $h^0_i$   \> standard specific enthalpy of the ice water ($298$~kJ~kg${}^{-1}$) \\
 $h^{\star}_v$ \> a specific moist static energy \citep{ambaum10} \\
 $k$      \> a dummy index for $(d,v,l,i)$ (for instance in $b_k$)\\
 $k_w, k$      \> liquid-water and  moist enthalpy in \citet{emanuel94} \\
 $k_{RD}$      \> the Randall-Deardorf CTEI parameter in the ($q_t$, $\theta_e$)  diagram\\
 $k_L$      \> the  Lilly CTEI parameter in the  ($q_t$, $\theta_l$)  diagram\\
 $K_{\psi}$      \> a general exchange coefficient for any  ${\psi}$\\
 $K_h$      \> the exchange coefficient for heat \\
 $K_q$      \> the exchange coefficient for water content \\
 $K_s$      \> the exchange coefficient for entropy \\
 ${\Lambda}_r$ \> $= [ (s_{v})_r - (s_{d})_r ] / c_{pd} \approx 5.87$ \\
 ${\Lambda}_0$ \> $= [ s_{v}^0 - s_{d}^0 ] / c_{pd} \approx 3.53$ \\
 $L_{\rm vap}$    \> $=h_v-h_l$: Latent heat of vaporisation \\
 $L^0_{\rm vap}$  \> $= 2.501 \times 10^{6}$~J~kg${}^{-1}$ at $T_0$\\
 $L_{\rm fus} $   \> $=h_l-h_i$: Latent heat of fusion \\
 $L^0_{\rm fus} $ \> $= 0.334 \times 10^{6}$~J~kg${}^{-1}$ at $T_0$ \\
 $L_{\rm sub}$    \> $=h_v-h_i$: Latent heat of sublimation \\
 $L^0_{\rm sub}$  \> $= 2.835 \times 10^{6}$~J~kg${}^{-1}$ at $T_0$ \\
 $M(C)$ \>  a parameter in \citet{marquet_geleyn13} \\
 MSE \> Moist Static Energy (MSE$_d$, MSE$_m$, MSE$_l$, LIMSE, FMSE) \\
 $m$  \> a mass of moist air (and $m_d$, $m_v$, $m_l$, $m_i$) \\
 ${\mu}$ \>  moist-air Gibbs' function $h-T\:s$  (with moist-air $h$ and $s$) \\
${\mu}_d$ \>  dry-air Gibbs' function ($h_d-T\:s_d$) \\
${\mu}_v$ \>  water-vapour Gibbs' function ($h_v-T\:s_v$) \\
${\mu}_l$ \>  liquid-water Gibbs' function ($h_l-T\:s_l$) \\
${\mu}_l$ \> ice Gibbs' function ($h_i - T\:s_i$) \\
 $N^2$ \>  the squared Brunt-V\"ais\"al\"a frequency \\
 $N^2_{sw}$ \>  the liquid-water saturating  version of $N^2$ \\
 PV      \> the Potential Vorticity \\
 PV$_s$      \> the moist-air entropy Potential Vorticity \\
 $p$      \> ($=p_d + e$) the local value for the pressure \\
 $p_r$  \> ($=(p_d)_r + e_r$) the reference pressure ($p_r=p_0=1000$~hPa)\\
 $p_d$     \> local dry-air partial pressure \\
 $(p_d)_r$ \> reference dry-air partial pressure ($\equiv p_r-e_r =993.89$~hPa)\\
 $(p_d)_0$ \> standard dry-air partial pressure ($1000$~hPa) \\
 $p_0$     \> a conventional pressure ($1000$~hPa) \\
 $q_{d}$   \> $={\rho}_d / {\rho}$: specific content for dry air \\
 $q_{v}$   \> $={\rho}_v / {\rho}$: specific content for water vapour \\
 $q_{l}$   \> $={\rho}_l / {\rho}$: specific content for liquid water \\
 $q_{i}$   \> $={\rho}_i / {\rho}$: specific content for ice\\
 $q_t  $   \> $= q_v+q_l+q_i$: total specific content of water \\
 $r_{v}$   \> $=q_{v}/q_{d}$: mixing ratio for water vapour \\
 $r_{l}$   \> $=q_{l}/q_{d}$: mixing ratio for liquid water \\
 $r_{i}$   \> $=q_{i}/q_{d}$: mixing ratio for ice \\
 $r_{r}$   \> reference mixing ratio for water species, with\\
           \> $\eta\:r_{r} \equiv e_r / (p_d)_r$, 
             leading to $r_{r} \approx 3.82$~g~kg${}^{-1}$ \\
 $r_{sw}$     \> saturating water-vapour mixing ratio over liquid water  \\
 $r_{t}$   \> $=q_{t}/q_{d}$: mixing ratio for total water \\
 ${\rho}_d$   \> specific mass for the dry air  \\
 ${\rho}_v$   \> specific mass for the water vapour \\
 ${\rho}_l$   \> specific mass for the liquid water \\
 ${\rho}_i$   \> specific mass for ice \\
 ${\rho}$   \> specific mass for the moist air  \\
             \> $={\rho}_d+{\rho}_v+{\rho}_l+{\rho}_i$  \\
 $R_v$   \> water-vapour gas constant --\= ($461.52$~J~K${}^{-1}$~kg${}^{-1}$) \kill
 $R_d$   \> dry-air gas constant     \> ($287.06$~J~K${}^{-1}$~kg${}^{-1}$) \\
 $R_v$   \> water-vapour gas constant \> ($461.53$~J~K${}^{-1}$~kg${}^{-1}$) \\
 $R$     \> $ = q_d \: R_d + q_v \: R_v$: gas constant for moist air \\
 $R_i$     \> Richardson number ($N^2/S^2$) \\
 $\Hat{R}$   \> a  parameter similar to $C$ to mix the ``dry and wet limits''  \\
                     \> for turbulent values \citep{Lewellen2004}\\
 $s$       \> the moist-air specific entropy \\
 $s_{ref}$  \>  and $s_r$: reference values for entropy \\
 $s_{d}$  \>  entropy of dry air\\
 $s_{v}$  \>  entropy of water vapour\\
 $s_{l}$  \>  entropy of liquid water\\
 $s_{i}$  \>  entropy of ice\\
 $(s_{d})_r$  \>  reference value for dry-air  entropy ($6777$~J~K${}^{-1}$~kg${}^{-1}$) \\
 $(s_{v})_r$  \> reference value for water-vapour entropy ($12673$~J~K${}^{-1}$~kg${}^{-1}$) \\
 $s^0_d$   \> standard specific entropy for dry air ($6775$~J~K${}^{-1}$~kg${}^{-1}$) \\
 $s^0_v$   \> standard specific entropy for  water vapour ($10320$~J~K${}^{-1}$~kg${}^{-1}$) \\
 $s^0_l$   \> standard specific entropy for  liquid water  ($3517$~J~K${}^{-1}$~kg${}^{-1}$) \\
 $s^0_i$   \> standard specific entropy for ice ($2296$~J~K${}^{-1}$~kg${}^{-1}$) \\
 $S_e, S_l$  \> two moist-air entropies in \citet{Pauluis_al_10} \\
 $S_a$  \>  a weighted sum  of  $S_e$ and $S_l$ \citep{Pauluis_al_10}  \\
 $S^2$      \> the squared of the wind shear (in $R_i$) \\
 $t, dt$  \> time and time step \\
 $T$       \> local temperature \\
 $T_h$   \> the moist-air enthalpy temperature \\
 $T_r$   \> the reference temperature ($T_r\equiv T_0$) \\
 $T_{0}$   \> the zero Celsius temperature ($273.15$~K) \\
 $T_{\Upsilon}$    \> a constant temperature ($2362$~K) \\
 $\theta$              \> $ = T\:(p_0/p)^{\kappa}$: the dry-air potential temperature\\
 ${\theta}_{l}$     \> the liquid-water potential temperature \citep{Betts73} \\
 ${\theta}_{v}$   \> the virtual potential temperature  \\
 ${\theta}_{e}$   \> the equivalent potential temperature (companion to ${\theta}_{l}$)\\
 ${\theta}_{es}$  \> the saturating value of ${\theta}_{e}$ \\
 ${\theta}'_{e}$   \> the equivalent potential temperature deduced from ${\theta}'_w$\\
 ${\theta}'_w$      \> the pseudo-adiabatic wet-bulb potential temperature \\
 ${\theta}_{s}$    \> the moist-air entropy potential temperature \citep{marquet11}  \\
 ${\theta}_{sr}$   \> the reference value for ${\theta}_{s}$ \\
 $({\theta}_s)_1$ \> an approximate version for ${\theta}_{s}$ \\
 ${\theta}_{S}$   \> a moist-air potential temperature \citep{hauf_holler87}  \\
 ${\theta}^{\star}$    \> a moist-air potential temperature \citep{marquet93}  \\
 ${\Upsilon}(T)$ \> $= [ \: h_v(T) - h_d(T) \: ] / (c_{pd}\:T)$ \\
 ${\Upsilon}(T_r)$ \> $= [ \: (h_v)_r - (h_d)_r\:  ] / (c_{pd}\:T_r)$, ${\Upsilon}(T_0) \approx 9.5$\\
 $w$  \> the vertical component of the velocity  \\
 $z$    \> the vertical height  \\
 ------------------\=  ------------------------------------ --\= \kill
 ASTEX \> Atlantic Stratocumulus Transition Experiment \\
 BOMEX \> Barbados Oceanographic and Meteorological Experiment \\
 CTEI \> Cloud Top Entrainment Instability \\
 DYCOMS \> DYnamics and Chemistry Of Marine Stratocumulus\\
 FIRE \> First ISCCP Regional Experiment \\
 EPIC \> East Pacific Investigation of Climate \\
 EUCLIPSE \> European Union CLoud  Intercomparison, \\
                  \> Process Study and Evaluation project\\
 GCM \> General Circulation Model \\
 ISCCP \> International Satellite Cloud Climatology Project \\
 NWP \> Numerical Weather Prediction \\
 PBL \> Planetary Boundary Layer
\end{tabbing}

\bibliographystyle{ws-book-har}    
\bibliography{arXiv_thermo_v4}

\end{document}